\documentclass[10pt,journal,compsoc]{IEEEtran}

\usepackage[T1]{fontenc}
\usepackage[utf8]{inputenc}
\usepackage[american]{babel}
\usepackage{csquotes}
\usepackage{microtype}
\usepackage{paralist}
\setdefaultenum{(1)}{(a)}{(i)}{A.}
\usepackage{enumitem}
\usepackage{url}
\usepackage{flushend}
\usepackage{etoolbox}
\usepackage{fix-cm}
\usepackage{textpos}
\usepackage{mfirstuc}
\usepackage{titlecaps}

\usepackage{amsmath}
\usepackage{amssymb}
\usepackage{bm}
\usepackage{relsize}
\usepackage{siunitx}
\usepackage[super]{nth}

\usepackage{booktabs}
\usepackage{multirow}
\usepackage{colortbl}
\usepackage{makecell}
\usepackage{rotating}
\usepackage{threeparttable}

\usepackage{float}
\usepackage{graphicx}
\usepackage[table,dvipsnames]{xcolor}
\usepackage{subcaption}

\usepackage{listingsutf8}

\usepackage{xparse}
\usepackage{xspace}

\usepackage[xindy,acronym]{glossaries}

\usepackage[numbers,sort&compress]{natbib}

\usepackage[hidelinks,bookmarks=false]{hyperref}
\usepackage[all]{hypcap}

\usepackage[capitalise]{cleveref}

\newcommand{\vda}{$\hat{A}_{12}$}
\newcommand{\code}[1]{\texttt{#1}}
\newcommand{\tool}[1]{\textit{#1}}
\newcommand{\bigo}{\mathcal{O}}
\newcommand{\version}[1]{\texttt{#1}}

\newcommand{\numprojects}[0]{\num{10}}
\newcommand{\numversions}[0]{\num{161}}
\newcommand{\numbenchsdist}[0]{\num{1829}}

\newcommand{\numbenchparamsdist}[0]{\num{6460}}
\newcommand{\numbenchparamstotal}[0]{\num{59164}}

\newcommand{\numsas}{eight}
\newcommand{\numsoas}{two}
\newcommand{\nummoeas}{six}

\newcommand{\rqeffectiveness}{RQ1}
\newcommand{\rqalgos}{RQ1a}
\newcommand{\rqbaseline}{RQ1b}
\newcommand{\rqchangeawareness}{RQ1c}

\newcommand{\rqefficiency}{RQ2}

\newcommand{\project}[1]{\textit{#1}}
\newcommand{\bytebuddy}{\project{Byte~Buddy}}
\newcommand{\eclipsecollections}{\project{Eclipse~Collections}}
\newcommand{\jctools}{\project{JCTools}}
\newcommand{\jenetics}{\project{Jenetics}}
\newcommand{\logging}{\project{Log4j~2}}
\newcommand{\netty}{\project{Netty}}
\newcommand{\okio}{\project{Okio}}
\newcommand{\rxjava}{\project{RxJava}}
\newcommand{\xodus}{\project{Xodus}}
\newcommand{\zipkin}{\project{Zipkin}}

\newcommand{\metric}[1]{\textit{#1}}
\newcommand{\apfd}{\metric{APFD}}
\newcommand{\apfdp}{\metric{APFD-P}}

\newcommand{\topthree}{\metric{Top-3}}

\newcommand{\priostrategy}[1]{\textit{#1}}

\newcommand{\prioadditional}{\priostrategy{Additional}}
\newcommand{\priototal}{\priostrategy{Total}}
\newcommand{\covca}[1]{\textit{#1}}
\newcommand{\covcap}{\covca{cac}}
\newcommand{\covcas}{\covca{car}}
\newcommand{\covnca}{\covca{nca}}

\newcommand{\sa}[1]{\textit{#1}}
\newcommand{\saGreedy}{\sa{Greedy}}
\newcommand{\saHC}{\sa{HC}}
\newcommand{\saGA}{\sa{GA}}
\newcommand{\saIBEA}{\sa{IBEA}}
\newcommand{\saMOCell}{\sa{MOCell}}
\newcommand{\saNSGAII}{\sa{NSGAII}}
\newcommand{\saNSGAIII}{\sa{NSGAIII}}
\newcommand{\saPAES}{\sa{PAES}}
\newcommand{\saSPEA}{\sa{SPEA2}}

\newcommand{\objs}[1]{\textit{#1}}
\newcommand{\objsc}{\objs{C}}
\newcommand{\objsco}{\objs{CO}}
\newcommand{\objsch}{\objs{CH}}

\newcommand{\objsccoch}{\objs{C-CO-CH}}

\newcommand{\nrcols}[2]{\multicolumn{#1}{l}{#2}}

\definecolor{greylight}{RGB}{240,240,240}
\definecolor{greymedium}{RGB}{189,189,189}
\definecolor{greydark}{RGB}{99,99,99}

\usepackage{tcolorbox}
\tcbset{colback=greylight, colframe=greydark, arc=1pt, boxrule=0.75pt}
\newcommand{\summarybox}[2]{
	\begin{tcolorbox}
		\small{
			\textbf{#1:} #2
		}
	\end{tcolorbox}
}
 
\newacronym{api}{API}{application programming interface}
\newacronym{apbc}{APBC}{average percentage block coverage}
\newacronym{apdc}{APDC}{average percentage decision coverage}
\newacronym{apec}{APEC}{average percentage element coverage}
\newacronym{apfd}{\apfd}{average percentage of fault-detection}
\newacronym{apfdp}{\apfdp}{average percentage of fault-detection on performance}
\newacronym{apoe}{APOE}{average percentage overlapping elements}
\newacronym{apsc}{APSC}{average percentage statement coverage}
\newacronym{aptec}{APTEC}{average percentage total elements coverage}
\newacronym{auc}{AUC}{area under the curve}
\newacronym{cd}{CD}{continuous deployment}
\newacronym{cfa}{CFA}{control flow analysis}
\newacronym{cg}{CG}{call graph}
\newacronym{ci}{CI}{continuous integration}
\newacronym{cli}{CLI}{command line interface}
\newacronym{cpu}{CPU}{central processing unit}
\newacronym{db}{DB}{database}
\newacronym{esd}{ESD}{effect size difference}
\newacronym{ex3}{eX\textsuperscript{3}}{Experimental Infrastructure for Exploration of Exascale Computing}
\newacronym{hc}{\saHC}{Steepest Ascent Hill Climbing}
\newacronym{ga}{\saGA}{Generational Genetic Algorithm}
\newacronym{hpc}{HPC}{high-performance computing}
\newacronym{ibea}{\saIBEA}{Indicator-Based Evolutionary Algorithm}
\newacronym{io}{I/O}{input/output}
\newacronym{jar}{JAR}{Java Archive}
\newacronym{jdk}{JDK}{Java Development Kit}
\newacronym{jmh}{\tool{JMH}}{\tool{Java Microbenchmark Harness}}
\newacronym{jvm}{JVM}{Java Virtual Machine}
\newacronym{mad}{MAD}{median absolute deviation}
\newacronym{mocell}{\saMOCell}{Multi-Objective Cellular Genetic Algorithm}
\newacronym{moea}{MOEA}{multi-objective evolutionary algorithm}
\newacronym{maoea}{MaOEA}{many-objective evolutionary algorithm}
\newacronym{ndcg}{nDCG}{normalized discounted cumulative gain}
\newacronym{nsgaii}{\saNSGAII}{Non-Dominated Sorting Genetic Algorithm II}
\newacronym{nsgaiii}{\saNSGAIII}{Non-Dominated Sorting Genetic Algorithm III}
\newacronym{oss}{OSS}{open-source software}
\newacronym{paes}{\saPAES}{Pareto Archived Evolution Strategy}
\newacronym{pp}{pp}{percentage points}
\newacronym{rcn}{RCN}{The Research Council of Norway}
\newacronym{rq}{RQ}{research question}
\newacronym{rta}{RTA}{rapid type analysis}
\newacronym{rts}{RTS}{regression test selection}
\newacronym{sa}{SA}{search algorithm}
\newacronym{smbp}{SMBP}{software microbenchmark prioritization}
\newacronym{sbsmbp}{SBSMBP}{search-based software microbenchmark prioritization}
\newacronym{spe}{SPE}{software performance engineering}
\newacronym{spea2}{\saSPEA}{Strength Pareto Evolutionary Algorithm 2}
\newacronym{sut}{SUT}{software under test}
\newacronym{tcp}{TCP}{test case prioritization}
\newacronym{vcs}{VCS}{version control system}

\newcommand{\paragraphending}[2]{\textbf{#1#2}}
\renewcommand{\paragraph}[1]{\paragraphending{#1}{.}}
\newcommand{\paragraphquestion}[1]{\paragraphending{#1}{?}}

\begin{document}

\title{Evaluating Search-Based Software Microbenchmark Prioritization}

\author{
  Christoph~Laaber, Tao~Yue, and
  Shaukat~Ali\IEEEcompsocitemizethanks{\IEEEcompsocthanksitem C. Laaber and S. Ali are with the Simula Research Laboratory, Oslo, Norway\protect\\
    E-mail: \{laaber, shaukat\}@simula.no
    \IEEEcompsocthanksitem T. Yue is with the Beihang University, Beijing, China\protect\\
    E-mail: yuetao@buaa.edu.cn
  }}
 
\IEEEtitleabstractindextext{
\glsresetall{}
\begin{abstract}
    
    Ensuring that software performance does not degrade after a code change is paramount.
    A solution is to regularly execute software microbenchmarks, a performance testing technique similar to (functional) unit tests, which, however,
    often becomes infeasible due to extensive runtimes.
    To address that challenge, research has investigated regression testing techniques, such as \gls{tcp}, which reorder the execution within a microbenchmark suite to detect larger performance changes sooner.
    Such techniques are either designed for unit tests and perform sub-par on microbenchmarks or require complex performance models, drastically reducing their potential application.
    In this paper, we empirically evaluate single- and multi-objective search-based microbenchmark prioritization techniques to understand whether they are more effective and efficient than greedy, coverage-based techniques. 
    For this, we devise three search objectives, i.e., coverage to maximize, coverage overlap to minimize, and historical performance change detection to maximize.
    We find that \glspl{sa} are only competitive with but do not outperform the best greedy, coverage-based baselines.
    However, a simple greedy technique utilizing solely the performance change history (without coverage information) is equally or more effective than the best coverage-based techniques while being considerably more efficient, with a runtime overhead of less than \num{1}\%. These results show that simple, non-coverage-based techniques are a better fit for microbenchmarks than complex coverage-based techniques.\end{abstract}
 
}

\maketitle

\glsresetall{}

\section{Introduction}
Regression testing comprises effective techniques for revealing faults in continuously evolving software systems~\citep{harman:07}, e.g., as part of \gls{ci}~\citep{elbaum:14,liang:18,haghighatkhah:18,elsner:21}.
To detect performance problems, particularly, in software libraries and frameworks, microbenchmarks are the predominantly used performance testing technique, similar to unit tests for functional testing~\citep{stefan:17}.
However, their extensive runtimes inhibit their adoption in \gls{ci}~\citep{huang:14,stefan:17,laaber:18,laaber:20}.
In our previous work~\citep{laaber:20}, we found that \num{15}\% of the \gls{jmh} suites on GitHub run longer than three hours and \num{3}\% longer than \num{12} hours.
Therefore, it is inevitable to employ performance regression testing techniques, such as \gls{smbp} (\gls{tcp} on software microbenchmarks), to capture important performance changes as early as possible.

Catching performance problems early is crucial for industry.
Meta mentioned that performance regression testing is worth investigating~\citep{alshahwan:23}, and MongoDB has created a sophisticated solution to run benchmarks as part of \gls{ci}~\citep{daly:21}.
Though academic research on microbenchmarks has recently gained attention~\citep{leitner:17,stefan:17,laaber:18,costa:21,jangali:22}, performance regression testing on microbenchmark-level is still scarce, mostly focusing on \gls{rts} for performance tests~\citep{oliveira:17,alshoaibi:19,alshoaibi:22,chen:22}.
\citet{mostafa:17} are the first to apply \gls{smbp}, introducing a technique based on a complex performance-impact model, which is, however, only suited for collection-intensive software and non-trivial to apply.
In our previous work~\citep{laaber:21c}, we perform a large-scale study of coverage-based, greedy heuristics for \gls{smbp}, inspired by \gls{tcp}~\citep{rothermel:99,elbaum:01, luo:19}, and found that the studied techniques are not nearly as effective on microbenchmarks as on unit tests and impose a relatively large runtime overhead of \num{17}\%.

A promising approach is search-based techniques, which are highly effective for various software engineering optimization problems~\citep{achimugu2014systematic,zhang2020uncertainty,fraser:10,arcuri:19,li:07,islam:12,epitropakis:15,marchetto:16,dinucci:20}.
Motivated by these successes, in this paper, we define \gls{sbsmbp} problems and solve them with single-objective \glspl{sa} and \glspl{moea}, with (up to) three objectives:
\begin{inparaenum}
    \item coverage (\objsc{}),
    \item coverage overlap among benchmarks (\objsco{}),
    and
    \item historical performance change size (\objsch{}).
\end{inparaenum}
Moreover, we define new \saGreedy{} \gls{smbp} techniques based on each of the three objectives as well as their combination (\objsccoch{}).

We evaluate the \gls{sbsmbp} (\numsoas{} single-objective \glspl{sa} and \nummoeas{} \glspl{moea}) and \saGreedy{} \gls{smbp} techniques on \numprojects{} \gls{jmh} suites having \numbenchsdist{} distinct microbenchmarks with \numbenchparamsdist{} distinct parameterizations across \numversions{} versions, regarding
\begin{inparaenum}
    \item effectiveness measured by the \gls{apfdp};
    and
    \item efficiency measured as the runtime overhead.
\end{inparaenum}
The study compares the \gls{sbsmbp} and \saGreedy{} \gls{smbp} techniques to two greedy, coverage-based baselines, i.e., \priototal{} and \prioadditional{}~\citep{mostafa:17,laaber:21c}.

The results show that the best \gls{sbsmbp} technique, i.e., the single-objective \gls{ga} combining the three objectives with weighted-sum (\gls{ga} \objsccoch{}), performs competitively with the best greedy baseline, i.e., \priototal{}.
However, it does not outperform \priototal{} regarding effectiveness, with both having an overall median \gls{apfdp} of \num{0.60}, while only adding a minor additional overhead of \num{1}\% compared to below \num{1}\% for \priototal{} on top of the coverage extraction overhead.
The best \gls{moea} is \gls{mocell}, which exhibits a lower median \gls{apfdp} of \num{0.58} than \gls{ga} \objsccoch{} and \priototal{}.
Surprisingly, the \saGreedy{} technique that relies only on the historical performance change size (i.e., \saGreedy{} \objsch{}) has a higher median \gls{apfdp} of \num{0.65} than \priototal{}.
Statistically, however, the best \saGreedy{} and \gls{sbsmbp} techniques are only as effective as \priototal{}.
Per project, \saGreedy{} \objsch{} is more effective than \priototal{} for one project and never worse, while having a significantly lower overhead of \num{1}\% consistently across the studied projects, whereas the coverage-based techniques have overheads ranging between \num{8}\% and \num{105}\% across the projects and \num{17}\% on average.

These results reveal that 
the \saGreedy{} technique relying solely on the performance change history (\objsch{}) performs the best and is arguably the simplest to implement.
Hence, we recommend practitioners to employ the \saGreedy{} \objsch{} technique to achieve consistently earlier performance change detection, e.g., as part of \gls{ci}, and researchers to investigate non-coverage-based \gls{smbp} techniques in the future.

\vspace{0.5em}

\noindent
To summarize, the main contributions of this paper are:
\begin{itemize}
    \item the first paper to describe \gls{smbp} with search (\gls{sbsmbp});
    \item three search objectives (two of which are novel) to employ in \gls{smbp} algorithms;
    \item an experimental study showing the effectiveness and efficiency of the new \saGreedy{} \gls{smbp} and \gls{sbsmbp} techniques compared to greedy, coverage-based baselines;
    \item an experimental study on the impact of change-awareness on the \gls{smbp} techniques;
    and
    \item an experimental study on the effectiveness and efficiency of \numsoas{} single-objective \glspl{sa} and \nummoeas{} \glspl{moea}.
\end{itemize}
\noindent
We provide all data and scripts in our replication package~\citep{laaber:24:replication_package_v0_2_0} and the \glspl{smbp} implementations in \tool{bencher} \version{v0.4.0}~\citep{laaber:bencher-v0.4.0-zenodo}.
 
\section{Software Microbenchmarking with \glsentryshort{jmh}}
\label{sec:background:jmh}

Software microbenchmarking is a performance testing technique that can be seen as the equivalent of unit testing for functional testing.
A software microbenchmark, microbenchmark or benchmark for short and used thereafter, measures a performance metric, usually runtime or throughput, of small code units, such as statements and methods.

The \glsentryfull{jmh} is the de facto standard for defining and executing Java microbenchmarks.
They are defined in source code with annotations, similar to \tool{JUnit}.
\Cref{lst:jmh} shows a simplified example.
A benchmark is a method annotated with \code{@Benchmark} (lines \num{8}--\num{15}), which optionally takes parameters as input, i.e., an \code{Input} object containing an instance variable annotated with \code{@Param} (lines \num{9} and \num{17}--\num{21}), called the parameterization of a benchmark.

{

\definecolor{strings}{RGB}{0,128,0}
\definecolor{comments}{RGB}{128,128,128}
\definecolor{keywords}{RGB}{0,0,128}
\definecolor{annotations}{RGB}{128,128,0}
\definecolor{classfield}{RGB}{102,14,122}
\definecolor{methoddecl}{RGB}{0,0,0}
\definecolor{typeparam}{RGB}{32,153,157}
\definecolor{const}{RGB}{102,14,122}
\definecolor{num}{RGB}{0,0,255}

\lstset{ language=Java,
	stringstyle=\color{strings},
	commentstyle=\color{comments},
	moredelim=[is][\color{annotations}]{_an_}{_an_},
	moredelim=[is][\color{classfield}]{_cf_}{_cf_},
	moredelim=[is][\color{methoddecl}]{_md_}{_md_},
	moredelim=[is][\color{typeparam}]{_tp_}{_tp_},
	moredelim=[is][\color{const}]{_c_}{_c_},
	moredelim=[is][\color{num}]{_n_}{_n_},
	numberstyle=\tiny\color{black},
	showstringspaces=false,
	backgroundcolor=\color{white},				basicstyle=\fontsize{7}{8}\ttfamily,			breakatwhitespace=false,					breaklines=true,							captionpos=b,								extendedchars=true,							frame=single,								keepspaces=true,							keywordstyle=\color{keywords}\bfseries,		numbers=left,								numbersep=10pt,								rulecolor=\color{black},					showspaces=false,							showstringspaces=false,						showtabs=false,								tabsize=4,									title=\lstname,								stepnumber=1,
	firstnumber=1,
	xleftmargin=2em,
	xrightmargin=2em,
	framexbottommargin=5pt,
	framextopmargin=5pt,
	framexleftmargin=2em,
	framexrightmargin=2em,
}

\begin{lstlisting}[float=htbp, caption=Modified \gls{jmh} example from \rxjava{}, label=lst:jmh, belowskip=0pt]
_an_@Fork_an_(_n_3_n_)
_an_@Warmup_an_(iterations = _n_10_n_, time = _n_1_n_, timeUnit = TimeUnit._c_SECONDS_c_)
_an_@Measurement_an_(iterations = _n_20_n_, time = _n_1_n_, timeUnit = TimeUnit._c_SECONDS_c_)
_an_@BenchmarkMode_an_(Mode._c_SampleTime_c_)
_an_@OutputTimeUnit_an_(TimeUnit._c_NANOSECONDS_c_)
public class ComputationSchedulerPerf {

	_an_@Benchmark_an_
	public void _md_observeOn_md_(Input input) {
		LatchedObserver<Integer> o = input.newLatchedObserver();
		input.observable
			.observeOn(Schedulers.computation())
			.subscribe(o);
		o.latch.await();
	}

	_an_@State_an_(Scope._c_Thread_c_)
	public static class Input extends InputWithIncrementingInteger {
		_an_@Param_an_({ "1", "1000" })
		public int _cf_size_cf_;
	}
}
\end{lstlisting}

}
 
\begin{figure}[tbp]
    \centering
    \includegraphics[width=\columnwidth]{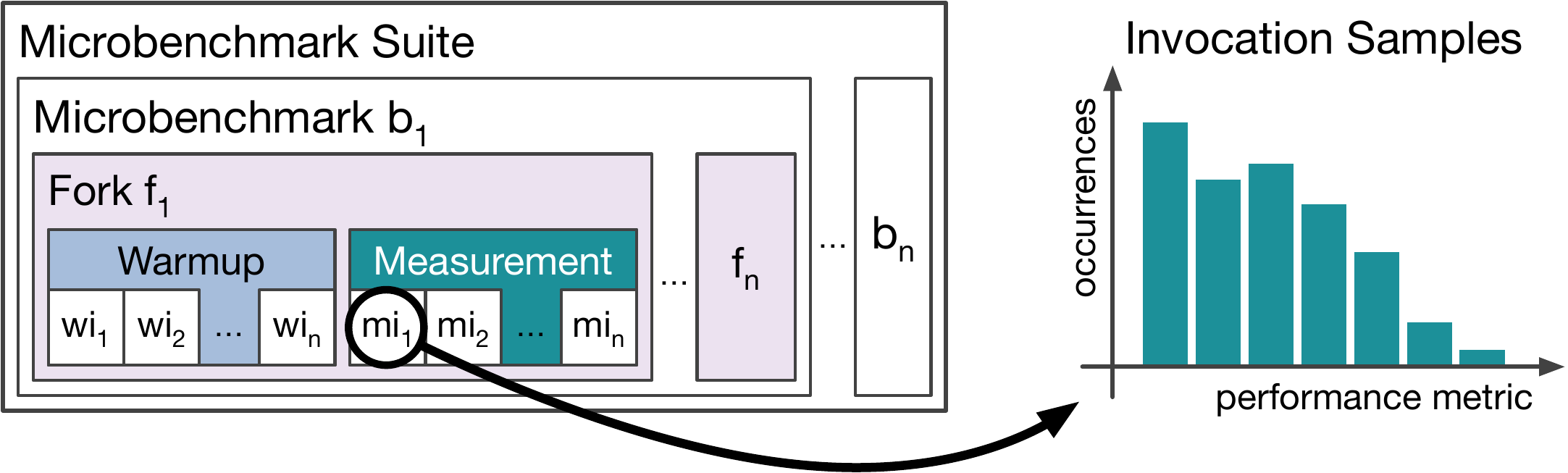}
    \caption{\Gls{jmh} Execution}
    \label{fig:jmh_exec}
\end{figure}

As measuring performance is non-deterministic and multiple factors can influence the measurement,
one has to repeatedly execute each benchmark to get reliable results.
\Gls{jmh} executes each combination of a benchmark method and its parameterization according to a specified configuration, set through annotations on class or method level (lines \num{1}--\num{5}) or through the \gls{cli}.
\Cref{fig:jmh_exec} visualizes the repeated execution of a benchmark suite.
Each (parameterized) benchmark is invoked as often as possible for a defined time period (e.g., \SI{1}{\second}, configured on lines \num{2} and \num{3}), called an iteration, and the measured performance metrics are reported.
To get reliable results, \gls{jmh} first runs a set of warmup iterations (line \num{2} and \enquote{wi}) to get the system into a steady state, for which the measurements are discarded.
After the warmup, \gls{jmh} runs a set of measurement iterations (line \num{3} and \enquote{mi}).
To deal with non-determinism of the \gls{jvm}, \gls{jmh} repeats the sets of warmup and measurement iterations for a number of forks (line \num{1} and \enquote{f}), each in a new \gls{jvm} instance.
The result of a benchmark is then the distribution of all measurement iterations (\enquote{mi}) of all forks (\enquote{f}).
A microbenchmark suite usually contains many benchmarks, which \gls{jmh} executes sequentially.

\section{Search-Based Prioritization}
\label{sec:approach}

This section defines \gls{sbsmbp}, which draws inspiration from search-based \gls{tcp}~\citep{li:07,islam:12,epitropakis:15,marchetto:16,dinucci:20}  and greedy \gls{smbp}~\citep{laaber:21c}.

\begin{figure*}[tbp]
    \centering
    \includegraphics[width=\textwidth]{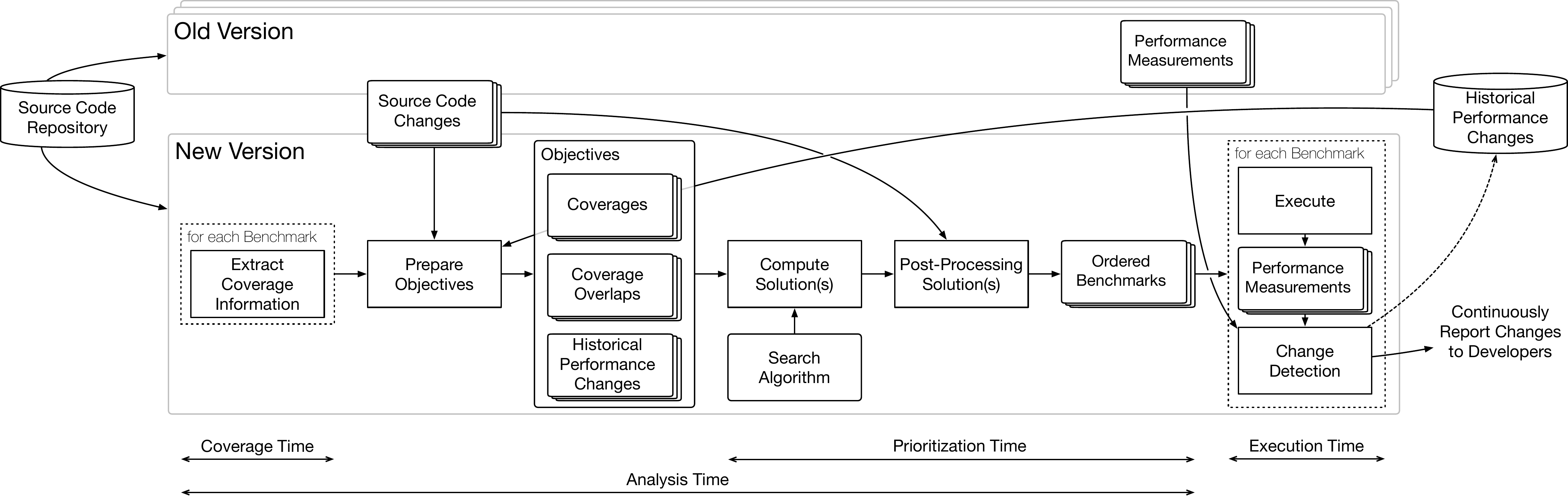}
    \caption{\titlecap{\glsentrylong{sbsmbp}} (adapted and extended from our previous work~\citep{laaber:21c})}
    \label{fig:approach}
\end{figure*}
 
\Cref{fig:approach} depicts an overview of \gls{sbsmbp}.
Upon a new version, the source code is retrieved from a repository.
First, the coverage information is extracted for every benchmark.
This is different from \gls{tcp} where coverage information is retrieved during the test execution, and the \gls{tcp} technique uses the coverage of the old version for ranking the test execution of the new version.
This is due to unit tests being usually only executed once.
Because benchmarks are executed repeatedly for rigorous measurements, \gls{sbsmbp} can leverage a single execution before the measurement to extract coverage information.
Based on the coverage information, historical performance changes, and source code changes (in case the technique implements change-awareness), a pre-processing stage prepares the search objectives.
Then, \gls{sbsmbp} employs a \gls{sa} to compute a \gls{smbp} ranking (solution), i.e., ordered benchmarks.
\Gls{sbsmbp} can be parameterized with either a single-, multi-, or many-objective \gls{sa}.

The post-processing stage has two tasks.
\begin{inparaenum}
    \item Select one solution from the \gls{sa} results.
    In the single-objective \gls{sa} case there is only one solution.
    In the multi-objective \gls{sa} case, the \gls{sa} returns the Pareto front containing multiple solutions to select from~\citep{fonseca:95}.
    \item Adjust the benchmark order based on the source code changes, when using a change-aware technique (see \cref{sec:study:indep-vars:ca}).
\end{inparaenum}
Finally, \gls{sbsmbp} executes the whole benchmark suite in the optimized order.
That is, it executes each benchmark repeatedly according to performance engineering best practice~\citep{georges:07}, measures the performance, and compares the measurements to the old version for change detection.
The changes are then reported to the developers and stored for future versions, i.e., to compute the historical performance changes objective.

\section{Experimental Study}

We conduct a laboratory experiment~\citep{stol:18} to empirically study the effectiveness and efficiency of \gls{sbsmbp}.

\subsection{Research Questions}

We pose the following \glspl{rq}:

\begin{description}[topsep=0.5em,labelindent=0.5em,labelwidth=2em,itemsep=0.5em]
    \item[\rqeffectiveness] How effective is \gls{sbsmbp}?
    \begin{description}[topsep=0.5em,labelwidth=2em,itemsep=0.2em]
        \item[\rqalgos] Which \gls{sbsmbp} technique is most effective?
        \item[\rqbaseline] How do \gls{sbsmbp} techniques compare to greedy techniques? 
        \item[\rqchangeawareness] How does change-awareness impact effectiveness?
    \end{description}    
    \item[\rqefficiency] How efficient is \gls{sbsmbp}? 
\end{description}

\noindent
\rqeffectiveness{} addresses the effectiveness of \gls{sbsmbp} by first investigating different \glspl{sa} and objectives in \rqalgos{} to find the most effective \gls{sbsmbp} techniques,
which we subsequently compare with greedy \gls{smbp} techniques in \rqbaseline{}, including the \priototal{} and \prioadditional{} baselines.
\rqchangeawareness{} assesses the impact of two source code change-aware techniques compared to the non-change-aware technique on \gls{smbp} effectiveness.
Finally, \rqefficiency{} studies the runtime overhead that the \gls{sbsmbp} techniques impose compared to the greedy techniques, which shows the practical feasibility.
 
\subsection{Dataset}
\label{sec:dataset}

To perform our experimental evaluation,
we require executions of
\begin{inparaenum}
    \item dedicated benchmarks (but not unit tests utilized as performance tests because their execution is not rigorous concerning performance testing and benchmarking best practice~\citep{georges:07});
    \item full benchmark suites, as \gls{tcp}/\gls{smbp} is defined by executing a full suite in a certain order~\citep{rothermel:99,rothermel:01};
    \item multiple versions of the same project to perform regression testing;
    and
    \item multiple projects to improve external validity.
\end{inparaenum}
In addition, a rigorous performance evaluation demands high standards and compliance with the performance engineering best practice to reduce internal validity threats~\citep{georges:07,maricq:18}.
We are only aware of one dataset that adheres to these criteria, which is from our previous work and was specifically created for \gls{smbp}~\citep{laaber:21c,laaber:21c:replication_package}.
The dataset includes
\begin{inparaenum}
    \item \gls{jmh} benchmark suite executions,
    \item dynamic coverage information on method level for these benchmarks,
    and
    \item coverage-based, greedy baselines for \gls{tcp} on benchmarks.
\end{inparaenum}
Though there are a few papers on microbenchmarking and performance regression testing~\citep{chen:17,mostafa:17,oliveira:17,chen:22,traini:22b},
no other dataset fits the above-mentioned criteria.
Hence, we only select the dataset of our previous work~\citep{laaber:21c,laaber:21c:replication_package} for our experiment.

\subsubsection{Study Objects}
\label{sec:study-objects}

The dataset has \numprojects{} \gls{oss} Java projects hosted on GitHub, which contain \gls{jmh} suites.
The projects have \numbenchparamstotal{} benchmarks in total and \numbenchparamsdist{} distinct benchmarks across \numversions{} versions.
Distinct benchmarks are counted once across all versions they occur in.
In this paper, we consider a benchmark to be the instantiation of a \gls{jmh} benchmark method (annotated with \code{@Benchmark}) with concrete parameterization (using \gls{jmh} parameters annotated with \code{@Param}).
\Cref{tab:study-objects} provides an overview of these projects.
Columns \enquote{Benchmarks} and \enquote{Runtime} show the arithmetic mean and standard deviation across the versions of each project, as different versions potentially have a different number of benchmarks and, hence, varying runtimes.

\begin{table}[tbp]
    \centering
    \footnotesize
    \caption{Study objects from the dataset~\citep{laaber:21c}}
    \label{tab:study-objects}

    \begin{tabular}{lrrrrr}
        \toprule
        Project & Versions & \nrcols{2}{Benchmarks} & \nrcols{2}{Runtime [h]} \\
        \cmidrule(l{1pt}r{1pt}){3-4}
        \cmidrule(l{1pt}r{1pt}){5-6}
        & & mean & stdev & mean & stdev \\
        \midrule

        \href{https://github.com/raphw/byte-buddy}{\bytebuddy{}} & \num{31} & \num{30.74} & $\pm$ \num{8} & \num{0.26} & $\pm$ \num{0.069} \\
        \href{https://github.com/eclipse/eclipse-collections}{\eclipsecollections{}} & \num{10} & \num{2371.40} & $\pm$ \num{13} & \num{38.45} & $\pm$ \num{0.124} \\
        \href{https://github.com/JCTools/JCTools}{\jctools{}} & \num{11} & \num{126.91} & $\pm$ \num{52} & \num{1.15} & $\pm$ \num{0.481} \\
        \href{https://github.com/jenetics/jenetics}{\jenetics{}} & \num{21} & \num{49.24} & $\pm$ \num{6} & \num{0.42} & $\pm$ \num{0.053} \\
        \href{https://github.com/apache/logging-log4j2}{\logging{}} & \num{15} & \num{309.53} & $\pm$ \num{162} & \num{2.71} & $\pm$ \num{1.398} \\
        \href{https://github.com/netty/netty}{\netty{}} & \num{10} & \num{746.50} & $\pm$ \num{522} & \num{6.56} & $\pm$ \num{4.625} \\
        \href{https://github.com/square/okio}{\okio{}} & \num{11} & \num{181.64} & $\pm$ \num{20} & \num{1.56} & $\pm$ \num{0.170} \\
        \href{https://github.com/ReactiveX/RxJava}{\rxjava{}} & \num{19} & \num{842.63} & $\pm$ \num{228} & \num{7.81} & $\pm$ \num{2.113} \\
        \href{https://github.com/JetBrains/xodus}{\xodus{}} & \num{11} & \num{67.00} & $\pm$ \num{10} & \num{1.33} & $\pm$ \num{0.104} \\
        \href{https://github.com/openzipkin/zipkin}{\zipkin{}} & \num{22} & \num{55.18} & $\pm$ \num{11} & \num{0.48} & $\pm$ \num{0.101} \\

        \bottomrule
    \end{tabular}
\end{table}

\subsubsection{Benchmark Executions and Performance Changes}
\label{sec:executions-changes}

We executed the benchmarks in a controlled environment for a predefined number of repetitions, following the best practice~\citep{georges:07,maricq:18}.
The executions used a unified configuration for all the benchmarks, i.e., \num{10} warmup iterations and \num{20} measurement iterations of \SI{1}{\second} each.
In addition, we executed the full suites for \num{3} trials (note that a trial is different from a \gls{jmh} fork, in that it is executed at different points in time and not back-to-back),
leading to a total runtime for all the projects, versions, and repetitions of approximately \num{89} days.

Based on the executions, the performance changes are computed between adjacent versions with a Monte-Carlo technique to estimate the confidence intervals of the ratio of the means.
The technique is based on \citet{kalibera:12,kalibera:13} and employs bootstrap~\citep{davison:97} with hierarchical random resampling with replacement~\citep{ren:10} on three levels, i.e., trial, iteration, and benchmark invocation.
It uses \num{10000} bootstrap iterations~\citep{hesterberg:15} and a confidence level of \num{99}\%.
For this, we used the \tool{pa} tool~\citep{laaber:pa-v0.1.0-zenodo}.

\subsubsection{Coverage Information and \glsentryshort{smbp} Baselines}
\label{sec:coverages-baselines}

The dataset provides the necessary coverage information for the \gls{sbsmbp}, \saGreedy{} \gls{smbp}, and greedy baseline techniques.
Dynamic coverage information, i.e., method coverage, is extracted per benchmark and version with \tool{JaCoCo}\footnote{\url{https://www.jacoco.org/jacoco/}}, by executing a benchmark once (with \gls{jmh}'s single-shot mode) and injecting the \tool{JaCoCo} agent.
Based on these coverages, we~\citep{laaber:21c} studied coverage-based, greedy \gls{smbp} techniques, particularly \priototal{} and \prioadditional{} --- our baselines.
 
\subsection{Independent Variables}
\label{sec:study:indep-vars}

Our experiment investigates four independent variables:
\begin{inparaenum}
    \item the prioritization strategies,
    \item the \glspl{sa} employed in the \gls{sbsmbp} techniques,
    \item the search objectives used for solving the \gls{sbsmbp} problem,
    and
    \item the change-awareness of the technique,
\end{inparaenum}
which are described below in detail.

\subsubsection{Prioritization Strategies}
\label{sec:study:indep-vars:strategy}

The four different prioritization strategies are:
\begin{inparaenum}
    \item the coverage-based, greedy \priototal{} strategy, which ranks the benchmarks in a suite, based on their total number of covered units, from the one with the most to the one with the least;
    \item the coverage-based, greedy \prioadditional{} strategy, which ranks the benchmarks, based on the number of covered units not yet been considered for prioritization by previously ranked benchmarks;
    \item a \saGreedy{} strategy which ranks the benchmarks according to either one or three of the search objectives;
    and
    \item the search-based strategy described in \cref{sec:approach}.
\end{inparaenum}
The strategies relying on code coverage use dynamic method coverage, aligned with our previous work~\citep{laaber:21c}.

The \priototal{} and \prioadditional{} strategies have their roots in unit testing~\citep{rothermel:99,rothermel:01}, are the standard coverage-based techniques with good effectiveness~\citep{hao:14,luo:19}, and were recently adapted for benchmarks~\citep{mostafa:17,laaber:21c}.
Our experiment parameterizes \priototal{} and \prioadditional{} with a benchmark granularity for both the prioritization strategy and the coverage extraction on the parameter level, which we showed to be optimal~\citep{laaber:21c}.
The \saGreedy{} strategy based on the search objectives uses either coverage, coverage-overlap, performance change history, or a combination of these.
The parameterization of the search-based strategy is mostly concerned with the employed \gls{sa} and search objectives, which are distinct independent variables of our experiment that the next sections describe.

\subsubsection{\titlecap{\glsentryfullpl{sa}}}
\label{sec:study:indep-vars:moea}
\label{sec:study:indep-vars:sa}

We study \numsas{} \glspl{sa}:
\begin{inparaenum}
    \item \glsentryfull{hc}, \glsunset{hc}
    \item \glsentryfull{ga}, \glsunset{ga}
    \item \glsentryfull{ibea}~\citep{zitzler:04} using the hypervolume~\citep{auger:09}, \glsunset{ibea}
    \item \glsentryfull{mocell}~\citep{nebro:09}, \glsunset{mocell}
    \item \glsentryfull{nsgaii}~\citep{deb:02}, \glsunset{nsgaii}
    \item \glsentryfull{nsgaiii}~\citep{deb:14}, \glsunset{nsgaiii}
    \item \glsentryfull{paes}~\citep{knowles:99}, \glsunset{paes}
    and
    \item \glsentryfull{spea2}~\citep{zitzler:01,kim:04}. \glsunset{spea2}
\end{inparaenum}
The first \numsoas{} are single-objective \glspl{sa}, and the other \nummoeas{} are \glspl{moea}.
We select these to cover a wide range of algorithms that have been used in previous search-based test optimization research~\citep{yoo:07,li:07,epitropakis:15,dinucci:20} and are supported by \tool{jMetal}~\citep{nebro:21}.

\paragraph{Solution Encoding}
A \gls{sbsmbp} solution is encoded as an integer permutation, where its length corresponds to the number of benchmarks in a suite, and each element corresponds to an integer identifier mapping to a unique benchmark in that suite.
We choose this encoding because the \gls{smbp} problem has the constraint that each benchmark exists exactly once in every solution.
\Gls{sbsmbp} then executes the benchmark suite in the order of the solution.

\paragraph{Algorithm Parameters}
Our experiment uses the same parameter settings across the \glspl{sa} where possible to facilitate a fair comparison, which are also in line with previous research on multi-objective \gls{tcp}~\citep{islam:12,li:13,epitropakis:15,marchetto:16,dinucci:20}.
The algorithm parameters are set to:
\begin{itemize}
    \item population size: \num{250};
    \item selection: binary tournament selection;
    \item crossover: PMX-Crossover with a probability $p_c = 0.9$;
    \item mutation: SWAP-Mutation with a probability $p_m = 1 / n$, where $n$ is the number of benchmarks to prioritize;
    \item maximum number of generations: \num{100};
    and
    \item maximum evaluations: population size times the maximum number of generations, i.e., \num{25000}.
\end{itemize}
Note that not all \glspl{sa} make use of all the parameters.
Archive-based \glspl{moea}, i.e., \gls{ibea}, \gls{mocell}, and \gls{paes}, use an archive size equal to the population size, which is in line with the papers that introduced them~\citep{zitzler:04,nebro:09,knowles:99} and the \tool{JMetal} defaults.
\Gls{mocell} is an exception, which requires a population size that is the power of \num{2} of an integer;
hence, we set \gls{mocell}'s population size to \num{256}, which is the closest to the population size of the other \glspl{moea}, i.e., \num{250}, and its maximum number of generations to \num{25600}.

\paragraph{\Gls{hc} Neighborhood}
We define the neighborhood of a solution as the benchmark orderings that swap the first benchmark with each of the other benchmarks.
The neighborhood size is $n-1$, where $n$ is the number of benchmarks in the suite.
This definition is in line with previous research to keep \gls{hc} scalable~\citep{li:07}.
Other definitions would also be valid, e.g., swapping any two benchmarks;
however, this would not scale, as for every iteration \gls{hc} would need to check $\bigo(n^2)$ neighbors.

\subsubsection{Search Objectives}
\label{sec:study:indep-vars:objectives}

The ideal search objective would be the actual performance changes (faults) the benchmark suite detects upon a new release, which, however, are only known after the execution (see \cref{sec:background:jmh} for details).
Hence, the search requires objectives that are good proxies for performance changes, which are available before the repeated suite execution.

Search-based \gls{tcp} often relies on coverage metrics~\citep{li:07,epitropakis:15,dinucci:20}, such as \gls{apec}~\citep{li:07} inspired by \gls{apfd}~\citep{li:07,dinucci:20}.
\gls{apec} requires $\bigo(mn)$, where $m$ and $n$ are the number of elements and tests, because it considers the additionally covered elements (\enquote{additional coverage}) for each test (not covered by an already ranked test)~\citep{epitropakis:15}, as it is a better proxy for fault detection than \enquote{total coverage}~\citep{luo:16a,luo:19}.

Differently for \gls{smbp}, \enquote{total coverage} leads to more effective rankings than \enquote{additional coverage}~\citep{laaber:21c}.
Hence, we refrain from using \gls{apec} and employ objectives that favor \enquote{total coverage}, i.e., \gls{aptec} in \cref{eq:aptec} inspired by \gls{apfdp}~\citep{mostafa:17,laaber:21c}.

\begin{equation}
    \label{eq:aptec}
    \textit{APTEC} = \frac{\sum\limits^{ n }_{ b=1 } \frac{ prev(b - 1) + elements(b) }{ m_{total} } }{ n }
\end{equation}
\noindent where $m_{total}$ is the number of total covered elements;
$n$ is the number of benchmarks;
$prev()$ is the cumulative element coverage until the previous benchmark $b-1$ in a \gls{smbp} ranking;
and
$elements(b)$ is the number of covered elements by benchmark $b$.
Note, $m_{total}$ potentially contains duplicate elements; e.g., if two benchmarks cover the same element, it is counted twice in $m_{total}$.
\gls{aptec} only requires $\bigo(n)$ to compute instead of $\bigo(mn)$ for \gls{apec}, as we can use memoization in $prev$: $prev(b) = prev(b-1) + elements(b)$, where $prev(0) = 0$.
What constitutes an element depends on the specific search objective.

We define \textbf{three objectives} based on \gls{aptec}:
\begin{description}
    \item [Coverage (\objsc{}, maximize):]
    This objective is akin to code coverage objectives in search-based \gls{tcp}~\citep{li:07}, but with \gls{aptec}, aiming to cover more code elements that are more likely to expose performance changes.
    While \citet{laaber:21c} showed that \gls{smbp} that solely relies on code coverage is only slightly more effective than a random strategy and code coverage has a low correlation with the performance change size, coverage is still one factor that can be used as a proxy for performance changes.
    Coverage maximizes the total number of code elements covered by a benchmark $b$ as returned by $elements(b)$.
    \item [Coverage Overlap (\objsco{}, minimize):]
    Due to Coverage greedily selecting benchmarks based on static coverage information (i.e., already covered code elements remain, which is akin to \emph{\priototal{}}), the \gls{sbsmbp} techniques are prone to rank benchmarks covering the same code elements similarly.
    To achieve coverage diversity among benchmarks, Coverage Overlap minimizes the cumulative coverage overlap among them.
    For Coverage Overlap, $elements(b)$ returns the average overlap between a benchmark $b$ and all other benchmarks in the suite.
    \item [Historical Performance Change (\objsch{}, maximize):]
    This objective maximizes the historically-detected performance change size of early-ranked benchmarks.
    The idea is that benchmarks that have previously detected large performance changes are more likely to detect performance changes in the future and, hence, should be ranked earlier.
    This is similar to search-based \gls{tcp}, which introduces \enquote{Fault History Coverage} as an objective~\citep{epitropakis:15}.
    For this objective, $elements(b)$ returns the average performance change size of benchmark $b$ across all previous versions.
\end{description}

Note, we do not employ delta coverage and benchmark execution cost as search objectives, as our preliminary experiments showed that delta coverage does not improve effectiveness and benchmark execution cost is approximately the same for all benchmarks when using a unified execution configuration (number of repetitions).

Our study investigates different combinations of \glspl{sa} and objectives across all \glspl{rq}. Since each objective can be considered a proxy metric for finding effective \gls{smbp} rankings, it is necessary to select one or more objectives to form different search problems and compare their effectiveness.
Particularly, with single-objective \glspl{sa}, we solve four problems:
\begin{inparaenum}
   \item three one-objective problems, each formed with each of the three objectives;
   and
   \item one three-objective problem, formed by aggregating the three objectives $o_i \in O$ into a single one $o^\prime$ with the classical weighted-sum approach, defined in \cref{eq:weighted-sum}, using equal weights $w_i = \frac{1}{3}$, similar to \citet{yoo:07}.
\end{inparaenum}
\begin{equation}
    \label{eq:weighted-sum}
    o^\prime = \sum_{i = 1}^{|O|} (w_i * o_i), \sum_{i = 1}^{|O|} w_i = 1
\end{equation}

\noindent We study the effectiveness and efficiency of the single-objective \glspl{sa} solving the above four problems and the \glspl{moea} solving the three-objective problem (with individual objectives).
We refer to the \gls{sbsmbp} techniques by its \gls{sa} and employed objectives, e.g., \gls{ga} \objsccoch{}.

\subsubsection{Change-Awareness}
\label{sec:study:indep-vars:ca}

This independent variable studies whether \enquote{change-awareness} of the \gls{smbp} technique improves its effectiveness.
That is, does the technique perform better if it considers the code that has changed since the last version in a dedicated way?
This is inspired by \citet{mostafa:17}, who studied change-aware approaches for \gls{smbp},
and
research on multi-objective search-based \gls{tcp} considering \enquote{delta coverage} as objective~\citep{epitropakis:15}.
We consider three approaches:
\begin{inparaenum}
    \item Non-Change-Aware (\covnca{}), which uses the full coverage information of the current version to encode the objectives \objsc{} and \objsco{};
    \item Change-Aware Coverage (\covcap{}), which relies only on coverage information that has changed since the last version, i.e., retains only changed methods, for the objectives \objsc{} and \objsco{};
    \item Change-Aware Ranking (\covcas{}), which uses the full coverage information (as for \covnca{}) to rank benchmarks and afterwards groups benchmarks covering a changed method before benchmarks not covering any changed method, while retaining the initial order.
\end{inparaenum}

\subsection{Dependent Variables}
\label{sec:study:dep-vars}

The dependent variables of our study involve a set of effectiveness and efficiency metrics: one effectiveness variable, i.e., \gls{apfdp}, to answer \rqeffectiveness{} and its sub-questions; and two efficiency variables, i.e., the prioritization and analysis (total) runtime overhead of the technique, to answer \rqefficiency{}.
They all have been used in previous research on \gls{smbp}~\citep{mostafa:17,laaber:21c}.

\subsubsection{Effectiveness}
\label{sec:study:dep-vars:effectiveness}

The effectiveness measures rely on the performance changes that the benchmarks of a suite detect between two adjacent versions.
We rely on the measurements and computed changes (see \cref{sec:executions-changes}) from our previous work~\citep{laaber:21c}, where a change is defined based on the bootstrapped confidence interval of the mean difference and not just the mean difference.
Note that we do not distinguish the two possible directions of a change, i.e., performance regression (slowdown) or improvement (speedup), but only consider these as a change of a certain size.
This aligns with both previous works on \gls{smbp}~\citep{mostafa:17,laaber:21c}.
Assuming that function $change$ returns the change of a benchmark between two versions as a percentage,
$change$ is defined as $change : B \mapsto \mathbb{Z}^{0+}$, where $B$ is the set of benchmarks in a suite.

\textbf{\Gls{apfdp}}~\citep{mostafa:17} adapts \gls{apfd}~\citep{rothermel:99} from unit testing.
\Gls{apfd} itself is inapplicable for benchmarks, as benchmarks have continuous outputs (i.e., different fault severities; see $change$) as opposed to the discrete ones (pass or fail) for unit tests.
\Gls{apfdp}, as an \gls{auc} metric, assesses the fault-detection capabilities of a \gls{smbp} technique, ranging from \num{0} to \num{1}, with \num{0} and \num{1} denoting the worst and optimal rankings of a benchmark suite, respectively.
A technique that ranks benchmarks with larger changes higher than those with smaller changes is considered better and has a larger \gls{apfdp} value, as defined in \cref{eq:apfdp}.
\begin{equation}
    \label{eq:apfdp}
    \textit{APFD-P} = \frac{\sum\limits^{ n }_{ x=1 } \frac{ detected(x) }{ c } }{ n }
\end{equation}
\noindent where $n$ is the benchmark suite size;
$c$ is the sum of the changes of all the benchmarks;
$detected(x)$ returns the cumulative change of the first $x$ benchmarks, see \cref{eq:detected}.
\begin{equation}
    \label{eq:detected}
    detected(x) = \sum\limits^{x}_{i=1} change(i)
\end{equation}
\noindent where $change(i)$ is the i\textsuperscript{th} benchmark's change in a ranking.

Previous \gls{smbp} works \citep{mostafa:17,laaber:21c} also study \gls{ndcg} and \topthree{}, which we do not consider due to similar results to \gls{apfdp} in our study.

\subsubsection{Efficiency}
\label{sec:study:dep-vars:efficiency}

Our experiment evaluates the technique's efficiency with two dependent variables in \rqefficiency{}.
Both concern the technique's runtime overhead imposed on the overall benchmarking time:
\begin{inparaenum}
    \item \textbf{Prioritization time} is the time required to run a \gls{smbp} technique, i.e., computing the \gls{smbp} ranking with all necessary inputs already available;
    and
    \item \textbf{Analysis time} is the total time necessary to prioritize a benchmark suite, including extracting all the required information (coverage information and historically-detected performance changes) and the prioritization time.
\end{inparaenum}
Both times are studied as overhead (in percentage) of the benchmark suite runtimes (see \cref{tab:study-objects}), which is required for comparability across the projects.
The overhead of \prioadditional{} and \priototal{} (this study's baselines) is dominated by the time to extract the coverage information~\citep{laaber:21c}.
Note that we neither study the coverage extraction time, as all the coverage-based techniques rely on the same coverage type,
nor the time to extract the historically detected changes, as these are available from previous versions and are not computed online at the time of prioritization.
 
\subsection{Experiment Setup}
\label{sec:study:setup}

To deal with the stochastic nature of the \gls{sbsmbp} techniques, we follow recommended practice~\citep{arcuri:11} and previous research on search-based \gls{tcp}~\citep{li:13,epitropakis:15,marchetto:16,dinucci:20},
and repeatedly execute each prioritization \num{30} times, i.e., for each project, version, and \gls{sa}, which are called \enquote{repetitions} in the rest of the paper.
The effectiveness calculations and results, unless otherwise specified, rely on all the repetitions for analyses.

Regarding the variance of a Pareto front (for each project, version, \gls{moea}, and repetition), we select the solution with the median effectiveness for analysis as similarly done by previous research on multi-objective \gls{tcp}~\citep{epitropakis:15,marchetto:16,dinucci:20}.
Other selection strategies, e.g, choosing the solution at a knee point~\citep{branke:04}, are arguably less applicable in our context where the objectives are just proxies for the context-dependent effectiveness (i.e., \gls{apfdp}) of \gls{smbp}.
The efficiency analysis, however, retains the measurements of all the solutions because more performance measurements raise the confidence that the results are representative~\citep{georges:07,maricq:18,laaber:19}.

Rigorously executing performance measurements is crucial to the validity of the results~\citep{georges:07,maricq:18}.
There are two basic principles to ensure reliable measurements:
\begin{inparaenum}
    \item sufficient numbers of repetitions
    and
    \item a controlled execution environment.
\end{inparaenum}
This paper assesses the overhead added to the full benchmark suite, which is in the order of minutes and hours (see \cref{tab:study-objects}).
Hence, minor measurement inaccuracies are permissible, as they will not change the overall results.
Best practice suggests \num{30} repeated measurements for every distinct measurement scenario (combinations of projects, versions, and \glspl{sa}).
We reuse the repetitions for tackling the \glspl{sa}['] stochasticity are as performance measurements, i.e., one measurement per repetition.

To ensure reliable measurements, a controlled execution environment is desirable.
However, due to the dimension of the runtimes, it is acceptable to use a slightly less controlled environment, as the results will not be impacted much.
Hence, we refrain from employing a tightly controlled bare-metal environment and use a \gls{hpc} cluster (\gls{ex3} cluster\footnote{\url{https://www.ex3.simula.no/}}) instead, as it allows for parallelization to keep the experiment runtime lower.
\gls{ex3} is hosted at the first author's institution, which uses Slurm \version{21.08.8-2} as its cluster management software.
The experiment executes the measurements on nodes of the same type (using the \tool{defq} partition of \gls{ex3}), with \num{30} \gls{cpu} cores assigned to and a single \gls{cpu} exclusively reserved for the measurement process.
The nodes have AMD EPYC 7601 \num{32}-core processors, run Ubuntu \version{22.04.3}, and have \SI{2}{\tera\byte} total memory shared for all \glspl{cpu}.
The experiments were conducted in the second half of 2023.
 
\subsection{Statistical Analyses}
\label{sec:statistical-analyses}

We employ hypothesis testing in combination with effect size measures to compare different observations.
An observation in our context can be, e.g., a single \gls{apfdp} value of a project, in a version, using one \gls{moea}, for a single repetition.
Depending on the analysis and the \gls{rq}, the compared set of observations changes, however, the tests remain the same.
We follow the best practice for evaluating search algorithms~\citep{arcuri:11}.

We use the non-parametric Kruskal–Wallis H~\citep{kruskal:52} test to compare multiple sets of observations.
Note that \citet{arcuri:11} suggest using the Mann-Whitney U test, which, however, only compares two sets of observations.
The Kruskal-Wallis H test is considered an extension of the Mann-Whitney U test and is, therefore, compatible with the best practice by \citet{arcuri:11}.
The null hypothesis H\textsubscript{0} states that the different distributions' medians are the same,
and the alternative hypothesis H\textsubscript{1} is that they are different.
If H\textsubscript{0} can be rejected, we apply Dunn's post-hoc test~\citep{dunn:64} to identify which pairs of observations are statistically different.

Note that we consciously decided against using the Friedman test, which is used by the SBFT\footnote{International Workshop on Search-Based and Fuzz Testing} tool competition~\citep{devroey:23} because our experiment setup \emph{does not} fit into a complete block design as required for the Friedman test.
To use the Friedman test (e.g., where the projects are the subjects and the \glspl{sa} are the treatments), we would need to aggregate the dependent variables of the different versions and experiment repetitions into a single value (e.g., taking the median of the median), which would completely disregard their variability.
Doing so is unacceptable and, hence, we use the Kruskal-Wallis H test instead, though it does not distinguish between the different subjects (projects).
To alleviate it, we add additional analysis on a per-project level.

We also report the effect size with Vargha-Delaney \vda{}~\citep{vargha:00} to characterize the magnitude of the differences.
Two groups of observations are the same if \vda{}~$ = 0.5$.
If \vda{}~$> 0.5$, the first group is larger than the second, otherwise if \vda{}~$< 0.5$.
The magnitude values are divided into four nominal categories, which rely on the scaled \vda{} defined
as $\hat{A}^{scaled}_{12} = (\hat{A}_{12} - 0.5) * 2$~\citep{hess:04}:
\enquote{negligible} ($|\hat{A}^{scaled}_{12}| < 0.147$),
\enquote{small} ($0.147 \leq |\hat{A}^{scaled}_{12}| < 0.33$),
\enquote{medium} ($0.33 \leq |\hat{A}^{scaled}_{12}| < 0.474$),
and
\enquote{large} ($|\hat{A}^{scaled}_{12}| \geq 0.474$).

All results are considered statistically significant at significance level $\alpha = 0.01$ and with a non-negligible effect size.
We control the false discovery rate with the Benjamini-Yekutieli procedure~\citep{benjamini:01} where multiple comparisons are performed.
It is considered a more powerful procedure than, e.g., the Bonferroni or Holm corrections for family-wise Type I errors.
 
\subsection{Threats to Validity}
\label{sec:threats}

\paragraph{Construct Validity}
We rely on \gls{apfdp} as the metric for \gls{smbp} effectiveness, which builds on the performance change size as a measure for benchmark importance.
Different metrics and measures are likely to impact the results and conclusions of this study.
Nevertheless, recent studies in performance regression testing have relied on the performance change size and \gls{apfdp} for \gls{smbp}~\citep{oliveira:17,mostafa:17,chen:21,laaber:21c}.
The procedure to compute the performance change size is from our previous work~\citep{laaber:21c}.
Consequently, this paper suffers from the same validity threats in this regard.

\paragraph{Internal Validity}
Regarding the threat introduced by the \glspl{sa}' stochasticity, we repeatedly executed the algorithms and chose the relevant statistical tests and the effect size measure by following the best practice reported by \citet{arcuri:11}.
Regarding the measurement inaccuracies of the runtime overhead,
we tried to mitigate this by executing on dedicated \gls{hpc} nodes and repeated runtime measurements, following the performance engineering best practices~\citep{georges:07}.
Regarding setting the \glspl{sa} hyperparameters, different hyperparameters might lead to different conclusions.
We followed previous research in search-based \gls{tcp}~\citep{dinucci:20} and used a unified configuration for all the \numsas{} \glspl{sa}.

\paragraph{External Validity}
The generalizability of our results is concerned with the number of datasets used, and consequently with the projects, versions, and benchmarks.
We rely on a single dataset from our previous work~\citep{laaber:21c}, which is, to the best of our knowledge, the only one that fits the study.All the projects are written in Java and use \gls{jmh} benchmarks.
Our results do not generalize to projects written in other languages and with other benchmarking frameworks.
The same limitation applies to other types of performance tests, such as application-level load and stress tests~\citep{jiang:15}, and macro- and application-benchmarks~\citep{daly:20,grambow:21}.
Finally, the results depend on performance changes executed in controlled, bare-metal environments;
hence, they are not generalizable to \gls{smbp} effectiveness assessed based on changes observed in less-controlled environments, e.g., the cloud.
  
\section{Results and Analyses}
\label{sec:results}

This section describes the results to answer the study's \glspl{rq}.

\subsection{\rqeffectiveness{}: Effectiveness}
\label{sec:results:effectiveness}

This section studies the effectiveness of \numsas{} different \glspl{sa} in \cref{sec:results:effectiveness:algos};
compares the best \glspl{sa} to the greedy baselines, i.e., \prioadditional{}, \priototal{}, and the \saGreedy{} strategy in \cref{sec:results:effectiveness:baselines};
and
assesses the impact of change-awareness in \cref{sec:results:effectiveness:change-awareness}.

\subsubsection{\rqalgos: Effectiveness of \Gls{sbsmbp} Techniques}
\label{sec:results:effectiveness:algos}

\begin{table*}[tbp]
    \centering

    \caption{
        \apfdp{} effectiveness for the \gls{sbsmbp} techniques (\glspl{sa} and objectives) and projects across all the versions and repetitions.
        The values are the median $\pm$ the \gls{mad}.
    }
    \label{tab:effectivness:algos}

    \fontsize{5.8pt}{7pt}\selectfont
    \begin{tabular}{llllllllllllll}
\toprule
\gls{sa} & Objectives & Overall & \multicolumn{10}{l}{Projects} \\
\cmidrule(l{1pt}r{1pt}){4-13}
& & & \bytebuddy{} & \eclipsecollections{} & \jctools{} & \jenetics{} & \logging{} & \netty{} & \okio{} & \rxjava{} & \xodus{} & \zipkin{} \\
\midrule

\saHC{} & \objsc{} & \num{0.506762}$\pm$\num{0.09932160} & \num{0.5172325}$\pm$\num{0.16047440} & \num{0.498355}$\pm$\num{0.02430426} & \num{0.4978665}$\pm$\num{0.07187274} & \num{0.527686}$\pm$\num{0.20048755} & \num{0.4998345}$\pm$\num{0.06776816} & \num{0.5107855}$\pm$\num{0.07929241} & \num{0.508015}$\pm$\num{0.13149698} & \num{0.500543}$\pm$\num{0.03757353} & \num{0.537514}$\pm$\num{0.09952768} & \num{0.511711}$\pm$\num{0.16552562} \\
 & \objsco{} & \num{0.5038235}$\pm$\num{0.09791535} & \num{0.5080125}$\pm$\num{0.16013933} & \num{0.4960975}$\pm$\num{0.02765642} & \num{0.500346}$\pm$\num{0.08278468} & \num{0.4905315}$\pm$\num{0.17164876} & \num{0.488936}$\pm$\num{0.05982958} & \num{0.5050625}$\pm$\num{0.07506256} & \num{0.559176}$\pm$\num{0.15964933} & \num{0.4989475}$\pm$\num{0.04112732} & \num{0.5070595}$\pm$\num{0.08339180} & \num{0.5229955}$\pm$\num{0.14854169} \\
 & \objsch{} & \num{0.508863}$\pm$\num{0.09828155} & \num{0.538462}$\pm$\num{0.16995711} & \num{0.5005655}$\pm$\num{0.02023082} & \num{0.5079715}$\pm$\num{0.06579186} & \num{0.525658}$\pm$\num{0.17846798} & \num{0.5022545}$\pm$\num{0.06393194} & \num{0.5068085}$\pm$\num{0.07982170} & \num{0.511251}$\pm$\num{0.13338656} & \num{0.5018895}$\pm$\num{0.04166106} & \num{0.5255545}$\pm$\num{0.10628240} & \num{0.513421}$\pm$\num{0.14456314} \\
 & \objsccoch{} & \num{0.508924}$\pm$\num{0.09593163} & \num{0.5377045}$\pm$\num{0.16068122} & \num{0.4997905}$\pm$\num{0.02948669} & \num{0.5087935}$\pm$\num{0.07527383} & \num{0.520743}$\pm$\num{0.18545992} & \num{0.505955}$\pm$\num{0.06703650} & \num{0.5111805}$\pm$\num{0.06720181} & \num{0.498731}$\pm$\num{0.13956900} & \num{0.4970275}$\pm$\num{0.03699828} & \num{0.500902}$\pm$\num{0.09551799} & \num{0.525249}$\pm$\num{0.14688934} \\

\saGA{} & \objsc{} & \num{0.563979}$\pm$\num{0.14424141} & \num{0.4902905}$\pm$\num{0.25786343} & \num{0.529408}$\pm$\num{0.02389284} & \num{0.6157785}$\pm$\num{0.10299103} & \num{0.62697}$\pm$\num{0.11269391} & \num{0.6036775}$\pm$\num{0.07584240} & \num{0.5234325}$\pm$\num{0.09434451} & \num{0.672824}$\pm$\num{0.10933211} & \num{0.5309465}$\pm$\num{0.06313874} & \num{0.697074}$\pm$\num{0.04593021} & \num{0.469462}$\pm$\num{0.16762720} \\
 & \objsco{} & \num{0.4811015}$\pm$\num{0.12520631} & \num{0.4}$\pm$\num{0.13691144} & \num{0.522739}$\pm$\num{0.03518210} & \num{0.4927835}$\pm$\num{0.07563262} & \num{0.383598}$\pm$\num{0.18134348} & \num{0.4353435}$\pm$\num{0.06846869} & \num{0.507442}$\pm$\num{0.08424207} & \num{0.7447565}$\pm$\num{0.15844398} & \num{0.513234}$\pm$\num{0.05188136} & \num{0.401718}$\pm$\num{0.05922839} & \num{0.5443955}$\pm$\num{0.12992172} \\
 & \objsch{} & \num{0.583442}$\pm$\num{0.12850584} & \num{0.675107}$\pm$\num{0.14200343} & \num{0.543922}$\pm$\num{0.02893739} & \num{0.592886}$\pm$\num{0.09239267} & \num{0.573101}$\pm$\num{0.27769098} & \num{0.5795235}$\pm$\num{0.07246801} & \num{0.5368515}$\pm$\num{0.08402191} & \num{0.588629}$\pm$\num{0.18009068} & \num{0.552105}$\pm$\num{0.07316631} & \num{0.5881945}$\pm$\num{0.14516359} & \num{0.5943285}$\pm$\num{0.13595071} \\
 & \objsccoch{} & \num{0.604286}$\pm$\num{0.12993655} & \num{0.6846975}$\pm$\num{0.13134502} & \num{0.5459325}$\pm$\num{0.02823834} & \num{0.6455615}$\pm$\num{0.07167259} & \num{0.5947055}$\pm$\num{0.20779529} & \num{0.613514}$\pm$\num{0.08676991} & \num{0.5384505}$\pm$\num{0.08808275} & \num{0.684705}$\pm$\num{0.13746890} & \num{0.5599185}$\pm$\num{0.07642581} & \num{0.709417}$\pm$\num{0.04564258} & \num{0.5503385}$\pm$\num{0.16169310} \\

\saIBEA{} & \objsccoch{} & \num{0.5700005}$\pm$\num{0.11730702} & \num{0.6130955}$\pm$\num{0.14566248} & \num{0.543536}$\pm$\num{0.03095002} & \num{0.5946255}$\pm$\num{0.06471623} & \num{0.577611}$\pm$\num{0.22776517} & \num{0.556529}$\pm$\num{0.06058682} & \num{0.51269975}$\pm$\num{0.05404114} & \num{0.689146}$\pm$\num{0.15420967} & \num{0.553465}$\pm$\num{0.05676023} & \num{0.6681035}$\pm$\num{0.07832353} & \num{0.5382175}$\pm$\num{0.14387373} \\

\saMOCell{} & \objsccoch{} & \num{0.577691}$\pm$\num{0.11929741} & \num{0.5929615}$\pm$\num{0.12711553} & \num{0.5604855}$\pm$\num{0.03490559} & \num{0.59627275}$\pm$\num{0.07057435} & \num{0.567075}$\pm$\num{0.20257431} & \num{0.56095}$\pm$\num{0.05995078} & \num{0.5298675}$\pm$\num{0.07241834} & \num{0.7630205}$\pm$\num{0.13285171} & \num{0.5779695}$\pm$\num{0.07374045} & \num{0.642716}$\pm$\num{0.07615857} & \num{0.51782875}$\pm$\num{0.17440787} \\

\saNSGAII{} & \objsccoch{} & \num{0.5611615}$\pm$\num{0.11098373} & \num{0.5951755}$\pm$\num{0.10979468} & \num{0.541484}$\pm$\num{0.01990020} & \num{0.592463}$\pm$\num{0.06337374} & \num{0.545322}$\pm$\num{0.19763873} & \num{0.553256}$\pm$\num{0.05479356} & \num{0.5167905}$\pm$\num{0.05213934} & \num{0.76739775}$\pm$\num{0.14290151} & \num{0.5508945}$\pm$\num{0.05518423} & \num{0.629235}$\pm$\num{0.07118482} & \num{0.5129565}$\pm$\num{0.16615869} \\

\saNSGAIII{} & \objsccoch{} & \num{0.5642005}$\pm$\num{0.11880407} & \num{0.59423625}$\pm$\num{0.14022616} & \num{0.5434795}$\pm$\num{0.02542659} & \num{0.58875275}$\pm$\num{0.06615917} & \num{0.57037}$\pm$\num{0.20037710} & \num{0.54628825}$\pm$\num{0.05727729} & \num{0.52847375}$\pm$\num{0.06592233} & \num{0.758318}$\pm$\num{0.14212092} & \num{0.552118}$\pm$\num{0.05814943} & \num{0.63372425}$\pm$\num{0.07843251} & \num{0.5176315}$\pm$\num{0.16966837} \\

\saPAES{} & \objsccoch{} & \num{0.523637}$\pm$\num{0.10385687} & \num{0.5428325}$\pm$\num{0.16833959} & \num{0.5018045}$\pm$\num{0.02594698} & \num{0.5573585}$\pm$\num{0.08064084} & \num{0.554503}$\pm$\num{0.17703727} & \num{0.5082215}$\pm$\num{0.06840272} & \num{0.5101445}$\pm$\num{0.06841309} & \num{0.577401}$\pm$\num{0.12694466} & \num{0.504467}$\pm$\num{0.03736597} & \num{0.568018}$\pm$\num{0.09083594} & \num{0.5293255}$\pm$\num{0.15395615} \\

\saSPEA{} & \objsccoch{} & \num{0.559295}$\pm$\num{0.11381216} & \num{0.58808775}$\pm$\num{0.11386553} & \num{0.5358605}$\pm$\num{0.01858624} & \num{0.59205125}$\pm$\num{0.05995709} & \num{0.5444}$\pm$\num{0.19075651} & \num{0.5501345}$\pm$\num{0.05048031} & \num{0.51571}$\pm$\num{0.04694542} & \num{0.7607145}$\pm$\num{0.16073979} & \num{0.548302}$\pm$\num{0.05367902} & \num{0.633113}$\pm$\num{0.07029192} & \num{0.5044815}$\pm$\num{0.18299621} \\

\bottomrule
\end{tabular}
 \end{table*}
 
This section compares the \gls{apfdp} effectiveness of the \numsas{} \glspl{sa} across all the projects and then per project.
In total, we study \num{14} \gls{sbsmbp} techniques, based on the \gls{sa} and search objective combinations (see \cref{sec:study:indep-vars:objectives}).
\Cref{tab:effectivness:algos} shows the effectiveness results overall and per project.

\begin{figure}[tbp]
    \centering
    \includegraphics[width=\columnwidth]{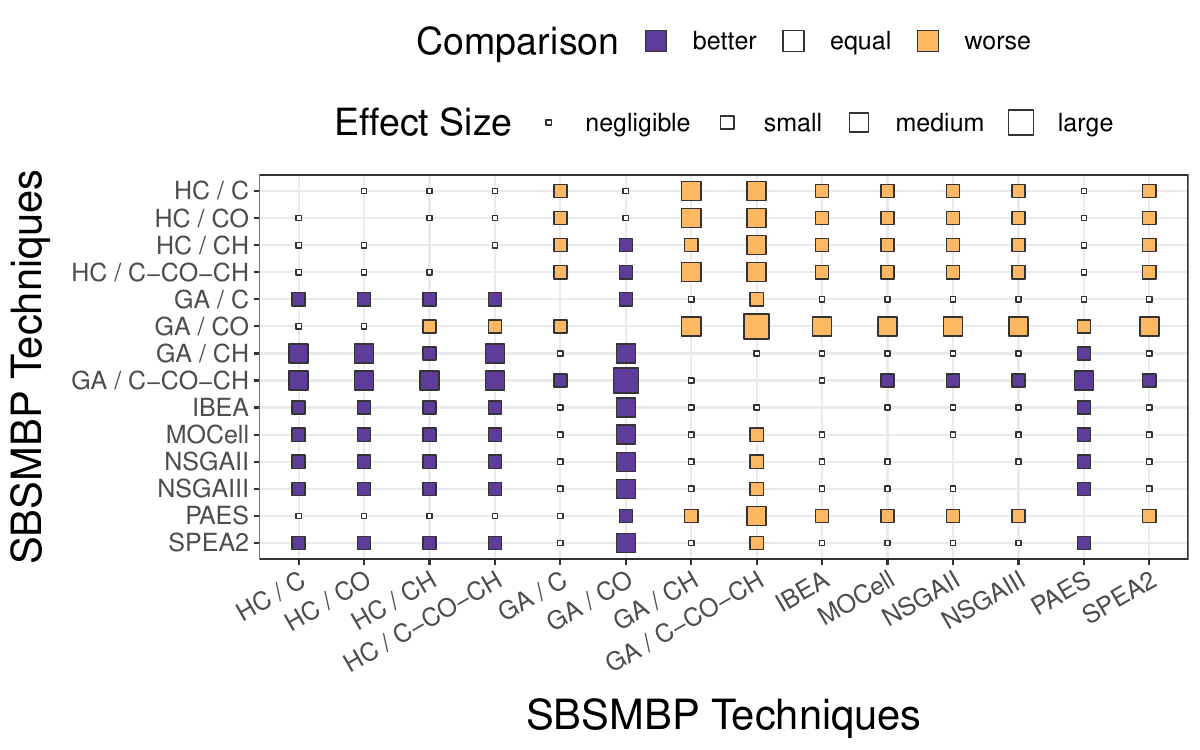}
    \caption{Overall \apfdp{} effectiveness compared pair-wise for the \gls{sbsmbp} techniques across all the versions and repetitions}
    \label{fig:effectiveness:stats:algos:overall}
\end{figure}
 
\paragraph{Overall}
From \cref{tab:effectivness:algos}'s \enquote{Overall} column, we observe that the median \gls{apfdp} ranges from \num{0.48} to \num{0.60}.
The \gls{sbsmbp} technique with the highest median \gls{apfdp} is \gls{ga} \objsccoch{} with \num{0.60}, followed by \gls{ga} \objsch{} with \num{0.58} and \gls{mocell} with \num{0.58}. Interestingly, \gls{ga} \objsch{} performs competitively by only relying on the historical performance change size and without relying on coverage.
We further notice that \gls{hc} (all objectives) and \gls{ga} \objsco{} exhibit the worst effectiveness.

In addition, we conduct pair-wise comparisons among all the \gls{sbsmbp} techniques with the statistical tests in \cref{fig:effectiveness:stats:algos:overall}.
The results confirm that \gls{hc} and \gls{ga} \objsco{} are statistically worse than all the other \glspl{sa} except \gls{paes}.
\Gls{hc} is the only local search algorithm and the most primitive in our study; consequently, it is unsurprising that it performs the worst.
\gls{ga} \objsco{} performs worse than the other \glspl{ga} and the \glspl{moea} because it only optimizes coverage overlap.

The single-objective \gls{ga} \objsccoch{} also performs the best statistically, better than all the other \glspl{sa} except \gls{ga} \objsch{} and \gls{ibea} with at least a small effect size.
This is interesting because a simple (single-objective) \gls{ga} using weighted-sum performs equally or better than the \glspl{moea}, which shows that in our context, using a \gls{moea} does not lead to higher effectiveness.
In addition, using \gls{ga} instead of any \gls{moea} has the benefit that the output is a single benchmark ranking and not a Pareto front from which a solution has to be selected.
The remaining \gls{ga} and \gls{moea} strategies are statistically equivalent though they show a minor difference in median effectiveness.

\begin{figure*}[tbp]
    \centering
    \begin{subfigure}{\textwidth}
        \includegraphics[width=\textwidth]{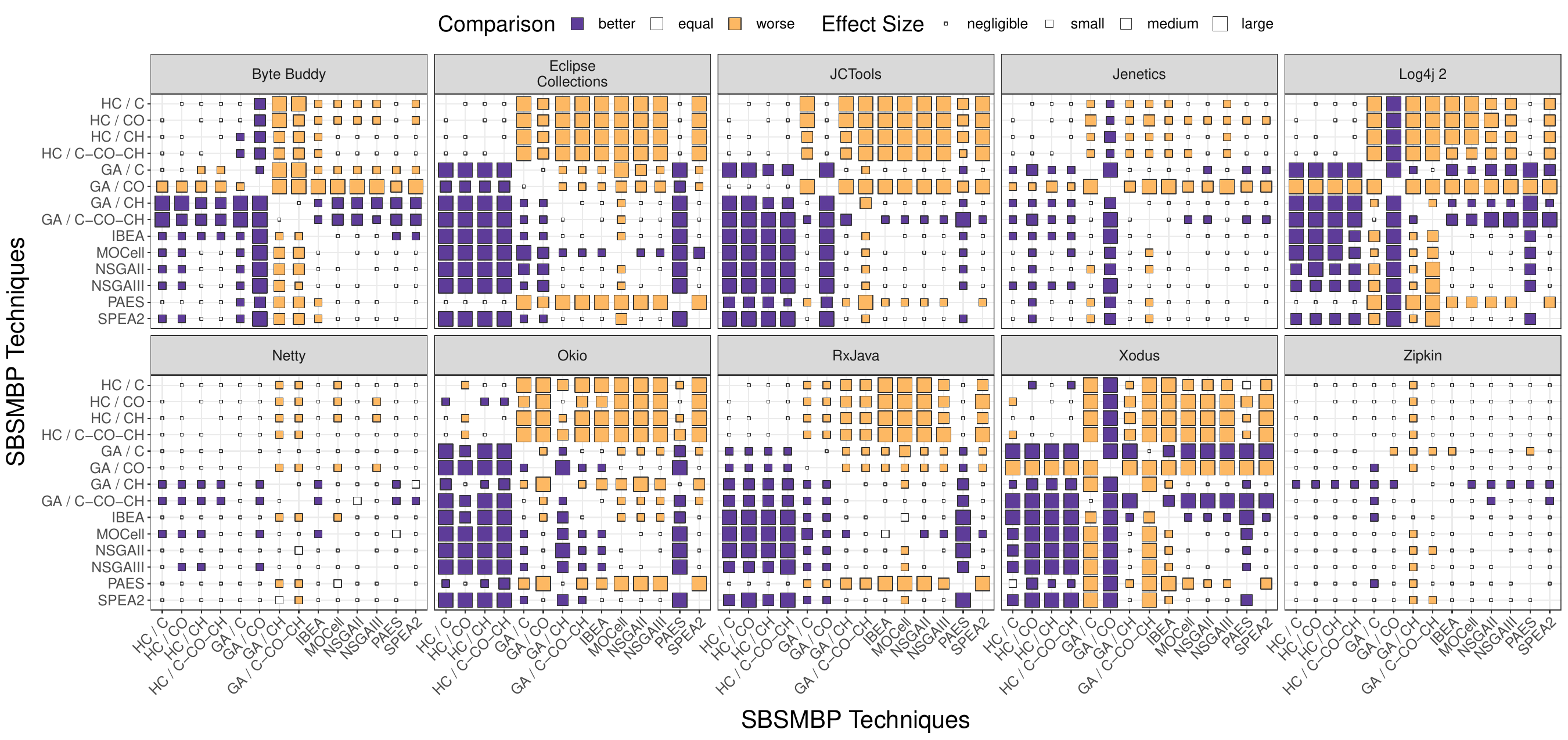}
    \end{subfigure}
    \caption{Per project \apfdp{} effectiveness compared pair-wise for the \gls{sbsmbp} techniques all the versions and repetitions}
    \label{fig:effectiveness:stats:algos:projects}
\end{figure*}
 
\paragraph{Per Project}
\Cref{tab:effectivness:algos} shows the median \gls{apfdp} per project, and \cref{fig:effectiveness:stats:algos:projects} depicts the pair-wise statistical test results.
We observe that the effectiveness depends on the project the \gls{smbp} technique is applied to, e.g., for \gls{ga} \objsccoch{}, \gls{apfdp} ranges from \num{0.54} (for \netty{}) to \num{0.71} (for \xodus{}).

We mostly notice a similar pattern to the overall results, i.e., \gls{hc} and \gls{ga} \objsco{} perform the worst.
The only exception is \okio{} where \gls{ga} \objsco{} performs well (\gls{nsgaii} performs even better on median, while the other \glspl{moea} except \gls{paes} are also effective).
The reason for this is that \okio{} is the project with the most disjoint coverage sets among the benchmarks and, consequently, a \gls{sbsmbp} technique with only the \objsco{} objective can perform well.
Moreover, we observe that for \netty{} and \zipkin{}, the \gls{sa} choice largely does not matter.

The most effective strategies for each project are:
\begin{inparaenum}
    \item \gls{ga} \objsccoch{} for six, i.e., \bytebuddy{}, \jctools{}, \jenetics{}, \logging{}, \netty{}, and \xodus{};
    \item \gls{ga} \objsch{} for two, i.e., \bytebuddy{} and \zipkin{};
    \item \gls{mocell} for two, i.e., \eclipsecollections{} and \rxjava{};
    and
    \item \gls{nsgaii} and \gls{ga} \objsco{} for one, i.e., \okio{}.
\end{inparaenum}

The reasons why a \gls{sa} is more effective depends on the characteristics of the studied projects and the problem definition.
As we are evaluating the \glspl{sa} on real-world projects, it is impossible to say which characteristics are responsible for the observed differences.
To provide such reasons, we would need to run a controlled experiment where we identify certain project characteristics, create projects and benchmark suites with these, and investigate their impact.
Such a study is out-of-scope of this paper.

\summarybox{\rqalgos{} Summary}{
    \Gls{ga} \objsccoch{} is the most effective \gls{sbsmbp} technique overall, as it performs the best for six of the ten projects.
    The other \gls{sbsmbp} techniques worth considering are \gls{ga} \objsch{}, particularly because it does not require coverage information, and \gls{mocell}, which is the most effective \gls{moea}.
}
 
\subsubsection{\rqbaseline{}: Comparison to Greedy Techniques}
\label{sec:results:effectiveness:baselines}

\begin{table*}[tbp]
    \centering

    \caption{
        \apfdp{} effectiveness for the greedy baselines, four \saGreedy{} techniques, and three best \gls{sbsmbp} techniques from \rqalgos{}; overall and per projects across all the versions and repetitions.
        The values are the median $\pm$ the \gls{mad}.
    }
    \label{tab:effectivness:baselines}

    \fontsize{5.5pt}{6.6pt}\selectfont
    \begin{tabular}{llllllllllllll}
\toprule
\gls{smbp} Algorithm & Objectives & Overall & \multicolumn{10}{l}{Projects} \\
\cmidrule(l{1pt}r{1pt}){4-13}
 & & & \bytebuddy{} & \eclipsecollections{} & \jctools{} & \jenetics{} & \logging{} & \netty{} & \okio{} & \rxjava{} & \xodus{} & \zipkin{} \\
\midrule

\priototal{} & -- & \num{0.602816}$\pm$\num{0.13293288} & \num{0.4896725}$\pm$\num{0.25424885} & \num{0.601873}$\pm$\num{0.03344301} & \num{0.6331815}$\pm$\num{0.12628861} & \num{0.650411}$\pm$\num{0.09931789} & \num{0.648952}$\pm$\num{0.07221374} & \num{0.586336}$\pm$\num{0.09473221} & \num{0.7045125}$\pm$\num{0.12551173} & \num{0.575914}$\pm$\num{0.10174120} & \num{0.693962}$\pm$\num{0.03612577} & \num{0.447917}$\pm$\num{0.13534359} \\

\prioadditional{} & -- & \num{0.517338}$\pm$\num{0.16609864} & \num{0.38419}$\pm$\num{0.16331061} & \num{0.603052}$\pm$\num{0.02898780} & \num{0.50157}$\pm$\num{0.04650471} & \num{0.44029}$\pm$\num{0.25686638} & \num{0.502365}$\pm$\num{0.06530779} & \num{0.50929}$\pm$\num{0.05222310} & \num{0.7402135}$\pm$\num{0.12734051} & \num{0.448204}$\pm$\num{0.13964016} & \num{0.6106185}$\pm$\num{0.11977036} & \num{0.581731}$\pm$\num{0.19562017} \\

\saGreedy{} & \objsc{} & \num{0.602667}$\pm$\num{0.13289285} & \num{0.4896725}$\pm$\num{0.25424885} & \num{0.598217}$\pm$\num{0.03606721} & \num{0.6358535}$\pm$\num{0.13016413} & \num{0.650411}$\pm$\num{0.09931789} & \num{0.648808}$\pm$\num{0.06976523} & \num{0.587018}$\pm$\num{0.09371070} & \num{0.7092965}$\pm$\num{0.12336047} & \num{0.5747775}$\pm$\num{0.09944836} & \num{0.693962}$\pm$\num{0.03612577} & \num{0.447917}$\pm$\num{0.13534359} \\
 & \objsco{} & \num{0.44399}$\pm$\num{0.16841743} & \num{0.3974495}$\pm$\num{0.15310365} & \num{0.587024}$\pm$\num{0.03910951} & \num{0.3871065}$\pm$\num{0.08675286} & \num{0.376569}$\pm$\num{0.15896882} & \num{0.3931485}$\pm$\num{0.07663930} & \num{0.424557}$\pm$\num{0.20268476} & \num{0.7971935}$\pm$\num{0.14006938} & \num{0.413014}$\pm$\num{0.14220284} & \num{0.3793575}$\pm$\num{0.05900081} & \num{0.555861}$\pm$\num{0.13533321} \\
 & \objsch{} & \num{0.648876}$\pm$\num{0.14731558} & \num{0.6727265}$\pm$\num{0.12932053} & \num{0.672696}$\pm$\num{0.04989246} & \num{0.5126545}$\pm$\num{0.08583216} & \num{0.621625}$\pm$\num{0.31684052} & \num{0.651328}$\pm$\num{0.06522773} & \num{0.484977}$\pm$\num{0.24299666} & \num{0.680838}$\pm$\num{0.17453390} & \num{0.636202}$\pm$\num{0.14441043} & \num{0.582346}$\pm$\num{0.15888802} & \num{0.631482}$\pm$\num{0.15675530} \\
 & \objsccoch{} & \num{0.606596}$\pm$\num{0.14209683} & \num{0.5002305}$\pm$\num{0.27572654} & \num{0.642749}$\pm$\num{0.03352752} & \num{0.651003}$\pm$\num{0.10423568} & \num{0.62029}$\pm$\num{0.13936514} & \num{0.6107625}$\pm$\num{0.09080777} & \num{0.597419}$\pm$\num{0.09438825} & \num{0.718441}$\pm$\num{0.15066255} & \num{0.5821585}$\pm$\num{0.11220984} & \num{0.718549}$\pm$\num{0.06115280} & \num{0.505342}$\pm$\num{0.17511730} \\

\saGA{} & \objsch{} & \num{0.583442}$\pm$\num{0.12850584} & \num{0.675107}$\pm$\num{0.14200343} & \num{0.543922}$\pm$\num{0.02893739} & \num{0.592886}$\pm$\num{0.09239267} & \num{0.573101}$\pm$\num{0.27769098} & \num{0.5795235}$\pm$\num{0.07246801} & \num{0.5368515}$\pm$\num{0.08402191} & \num{0.588629}$\pm$\num{0.18009068} & \num{0.552105}$\pm$\num{0.07316631} & \num{0.5881945}$\pm$\num{0.14516359} & \num{0.5943285}$\pm$\num{0.13595071} \\
 & \objsccoch{} & \num{0.604286}$\pm$\num{0.12993655} & \num{0.6846975}$\pm$\num{0.13134502} & \num{0.5459325}$\pm$\num{0.02823834} & \num{0.6455615}$\pm$\num{0.07167259} & \num{0.5947055}$\pm$\num{0.20779529} & \num{0.613514}$\pm$\num{0.08676991} & \num{0.5384505}$\pm$\num{0.08808275} & \num{0.684705}$\pm$\num{0.13746890} & \num{0.5599185}$\pm$\num{0.07642581} & \num{0.709417}$\pm$\num{0.04564258} & \num{0.5503385}$\pm$\num{0.16169310} \\

\saMOCell{} & \objsccoch{} & \num{0.577691}$\pm$\num{0.11929741} & \num{0.5929615}$\pm$\num{0.12711553} & \num{0.5604855}$\pm$\num{0.03490559} & \num{0.59627275}$\pm$\num{0.07057435} & \num{0.567075}$\pm$\num{0.20257431} & \num{0.56095}$\pm$\num{0.05995078} & \num{0.5298675}$\pm$\num{0.07241834} & \num{0.7630205}$\pm$\num{0.13285171} & \num{0.5779695}$\pm$\num{0.07374045} & \num{0.642716}$\pm$\num{0.07615857} & \num{0.51782875}$\pm$\num{0.17440787} \\

\bottomrule
\end{tabular}
 \end{table*}
 
This section compares the \gls{sbsmbp} techniques to the two greedy baselines (i.e., \prioadditional{} and \priototal{}) and four \saGreedy{} techniques, one for each objective \objsc{}, \objsco{}, and \objsch{} and the combination \objsccoch{} (see \cref{sec:study:indep-vars:objectives}).
For brevity, we only investigate the three best \gls{sbsmbp} techniques from \rqalgos{}, i.e., \gls{ga} \objsccoch{}, \gls{ga} \objsch{}, and \gls{mocell}.
\Cref{tab:effectivness:baselines} shows the effectiveness results overall and per project.

\begin{figure}[tbp]
    \centering
    \includegraphics[width=\columnwidth]{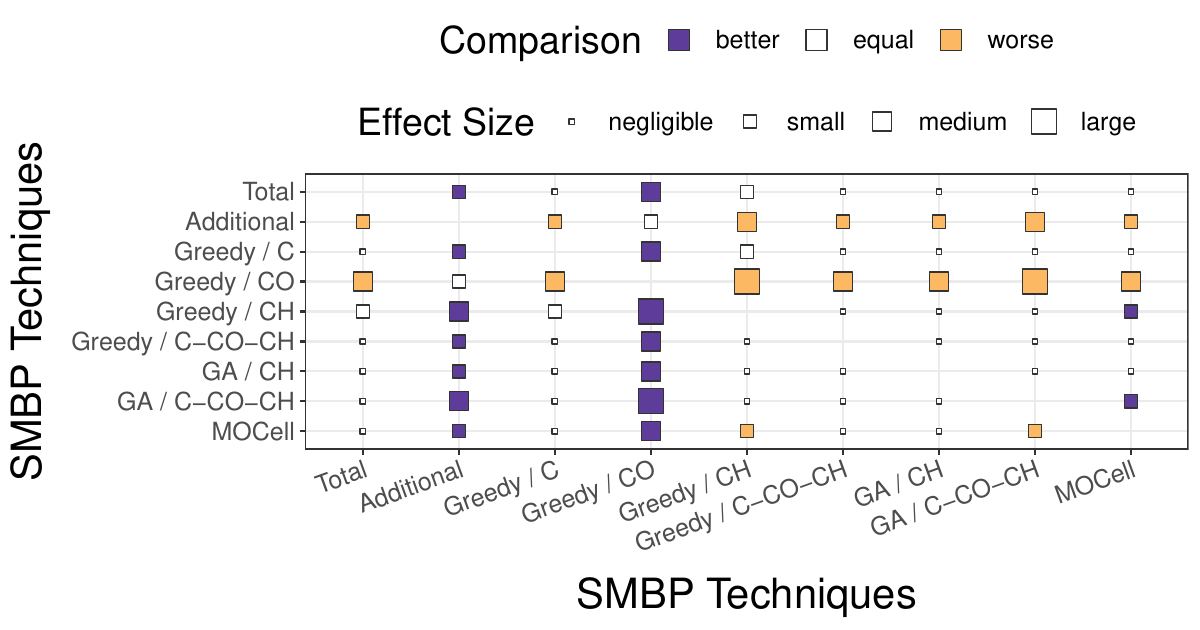}
    \caption{Overall \apfdp{} effectiveness pairwise comparison among the three best \gls{sbsmbp} technqiues from \rqalgos{}, the greedy baselines, and the \saGreedy{} techniques across all the versions and repetitions}
    \label{fig:effectiveness:stats:baselines:overall}
\end{figure}
 
\paragraph{Overall}
From \cref{tab:effectivness:baselines}'s column \enquote{Overall}, we observe that the median \gls{apfdp} ranges from \num{0.44} to \num{0.65}, where the two greedy baselines achieve \num{0.60} (for \priototal{}) and \num{0.52} (for \prioadditional{}).
We already found that \priototal{} is more effective than \prioadditional{} for \gls{smbp}~\citep{laaber:21c}, which might come as a surprise to \gls{tcp}-savvy readers (e.g., see \citet{luo:16a}).

A surprising result is that the \gls{sbsmbp} techniques exhibit either only an equal (\gls{ga} \objsccoch{}) or a lower (\gls{ga} \objsch{} and \gls{mocell}) median \gls{apfdp} compared to \priototal{}.
However, our observation aligns with \gls{tcp} research~\citep{luo:16a}, where the search-based technique is inferior to the greedy baseline (in their case \prioadditional{}).
The statistical tests, depicted in \cref{fig:effectiveness:stats:baselines:overall}, confirm that the \gls{sbsmbp} techniques are statistically equal to but never worse than \priototal{} and better than \prioadditional{} (with small and medium effect sizes).

The \saGreedy{} techniques show a similar pattern as the \gls{ga} techniques in \rqalgos{}: \saGreedy{} \objsco{} is the technique with the lowest median \gls{apfdp} of \num{0.44}, \saGreedy{} \objsc{} is equivalent to \priototal{} at \num{0.60}, and both \saGreedy{} \objsccoch{} with \num{0.61} and \saGreedy{} \objsch{} with \num{0.65} exhibit higher median \glspl{apfdp} than \priototal{}.
However, the improvement of the latter two \saGreedy{} techniques over \priototal{} is not statistically significant (see \cref{fig:effectiveness:stats:baselines:overall}).
Nevertheless, using \saGreedy{} \objsch{} and achieving a better median effectiveness even though being statistically equivalent is a surprising, yet great result, as it does not require coverage information.
\rqefficiency{} in \cref{sec:results:efficiency} investigates the implications of this on the technique efficiency.

\begin{figure*}[tbp]
    \centering
    \begin{subfigure}{\textwidth}
        \includegraphics[width=\textwidth]{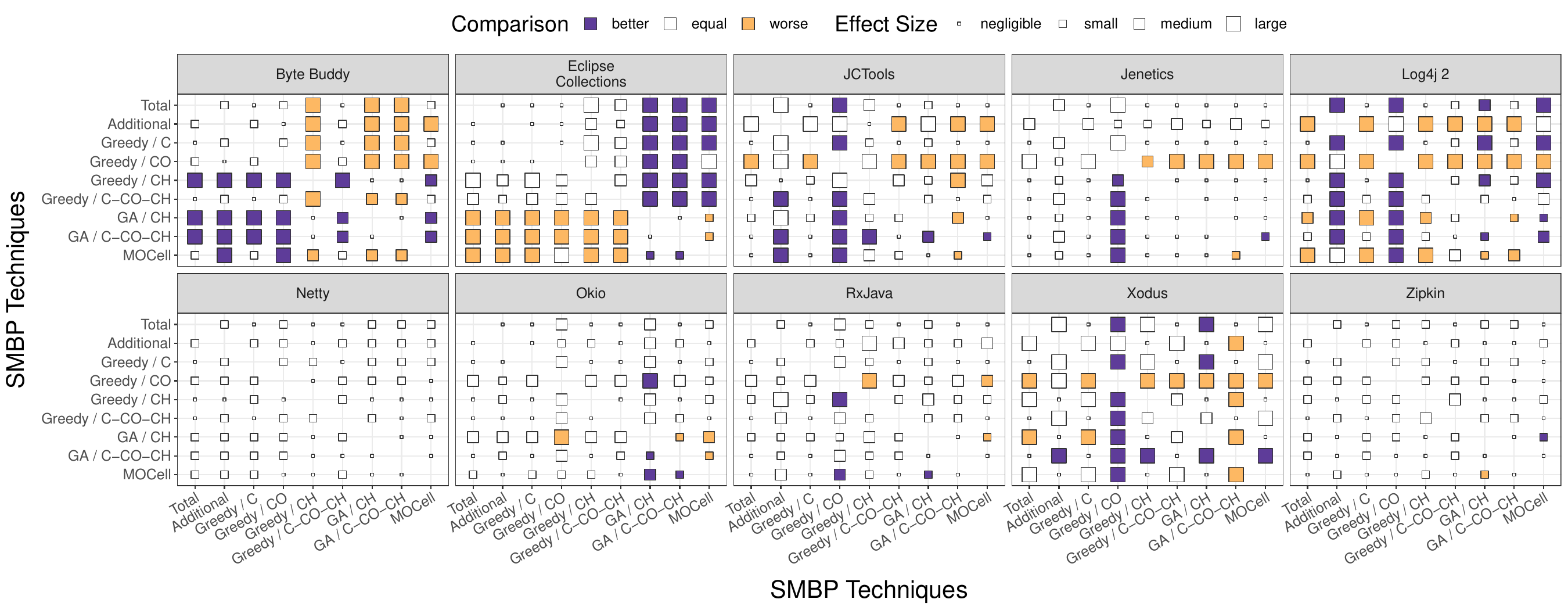}
    \end{subfigure}
    \caption{Per project \apfdp{} effectiveness pairwise comparison among the three best \gls{sbsmbp} technqiues from \rqalgos{}, the greedy baselines, and the \saGreedy{} techniques across all the versions and repetitions}
    \label{fig:effectiveness:stats:baselines:projects}
\end{figure*}
 
\paragraph{Per Project}
While overall the baseline \priototal{} is not statistically outperformed by any of the \gls{sbsmbp} and \saGreedy{} techniques, the per-project results offer a nuanced view in \cref{tab:effectivness:baselines,fig:effectiveness:stats:baselines:projects}.

The best \gls{sbsmbp} technique, i.e., \gls{ga} \objsccoch{} is statistically equivalent to \priototal{} for eight projects, better for one (\bytebuddy), and worse for one (\eclipsecollections{}).
The other two \gls{sbsmbp} techniques, i.e., \gls{ga} \objsch{} and \gls{mocell} perform worse.
\Gls{ga} \objsch{} is statistically better than \priototal{} for one project (\bytebuddy{}), worse for three, and equivalent for six,
whereas
\gls{mocell} is never statistically better, worse for two projects, and equivalent for eight.
The median differences are in line:
\begin{inparaenum}
    \item \gls{ga} \objsccoch{} is four times better and six times worse than \priototal{};
    \item \gls{ga} \objsch{} is twice better and eight times worse;
    and 
    \item \gls{mocell} is three times better, six times worse, and once equal.
\end{inparaenum}

The two best \saGreedy{} techniques perform favorably compared to \priototal{}:
\begin{inparaenum}
    \item \saGreedy{} \objsch{} is better for one project (\bytebuddy{}) and equal for nine, while the medians are better for four projects, worse for five, and equal for one;
    \item \saGreedy{} \objsccoch{} is equal for all the ten projects, while the medians are better for seven, worse for two, and equal for one. 
\end{inparaenum}

From a project perspective, the \gls{sbsmbp} techniques and \saGreedy{} \objsch{} perform better on \bytebuddy{}, while just the \gls{sbsmbp} techniques perform worse on \eclipsecollections{}.
\bytebuddy{} is the project with the smallest benchmark suite, and \eclipsecollections{} the one with the largest.
However, this suite size trend is not noticeable for the remaining projects.
For \jenetics{}, \netty{}, \okio{}, \rxjava{}, and \zipkin{}, it largely does not matter which technique one chooses as long as it is not \saGreedy{} \objsco{} and \gls{ga} \objsch{}.
For \jctools{}, \logging{}, and \xodus{} any technique except \prioadditional{}, \saGreedy{} \objsco{}, \gls{ga} \objsch{}, and \gls{mocell} is equally effective.

The strategies with the highest median \gls{apfdp} for each project are:
\begin{inparaenum}
    \item \saGreedy{} \objsch{} for four, i.e., \eclipsecollections{}, \logging{}, \rxjava{}, and \zipkin{};
    \item \saGreedy{} \objsccoch{} for three, i.e., \jctools{}, \netty{}, and \xodus{};
    \item \priototal{} for two, i.e., \jenetics{} and \logging{};
    \item \gls{ga} \objsccoch{} for two, i.e., \bytebuddy{} and \jctools{};
    and
    \item \gls{ga} \objsch{} for one, i.e., \bytebuddy{}.
\end{inparaenum}

\summarybox{\rqbaseline{} Summary}{
    Only one \gls{sbsmbp} technique, i.e., \gls{ga} \objsccoch{}, is competitive with the best greedy techniques.
    The baseline \priototal{} performs surprisingly well, only improved upon by \saGreedy{} \objsch{} for one project and when considering the median \gls{apfdp}.
    Considering that \saGreedy{} \objsch{} does not require coverage information makes it a great candidate to generally recommend.
}
 
\subsubsection{\rqchangeawareness{}: Impact of Change-Awareness}
\label{sec:results:effectiveness:change-awareness}

\begin{figure*}[tbp]
    \centering
    \includegraphics[width=\textwidth]{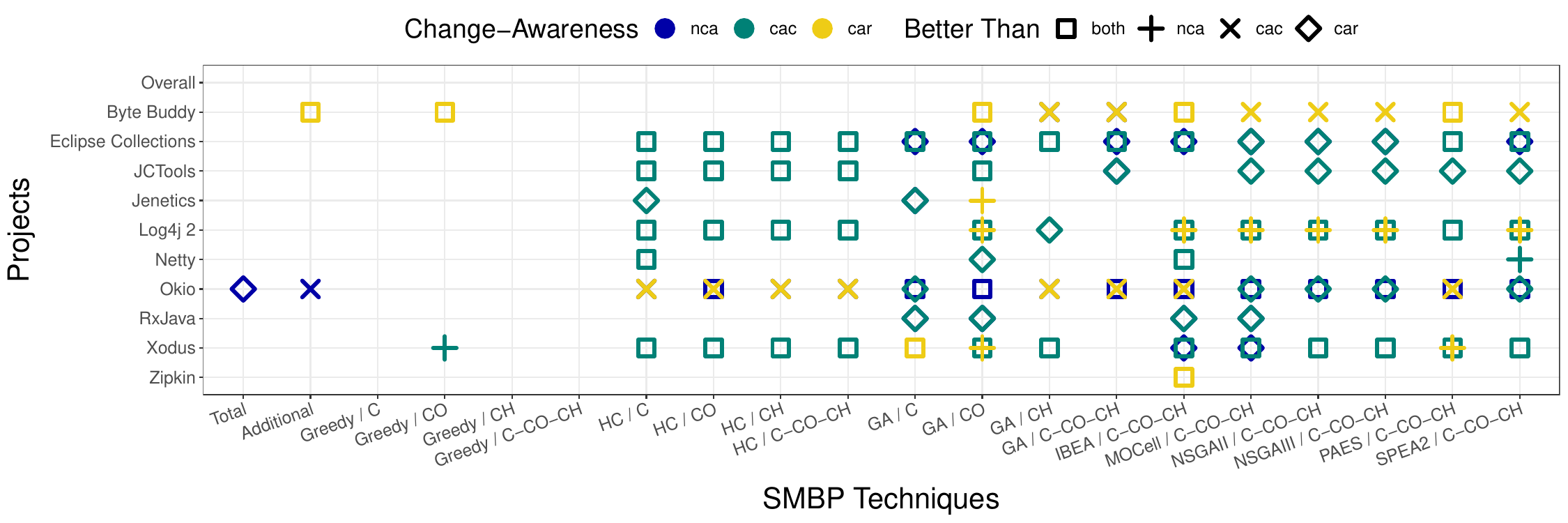}
    \caption{Per project change-awareness pair-wise comparison for each \gls{smbp} technique and project across all the versions and repetitions}
    \label{fig:effectiveness:change-awareness}
\end{figure*}
 
This section studies if the \gls{smbp} techniques from \rqalgos{} and \rqbaseline{} perform differently when incorporating change-awareness.
Specifically, we compare \covnca{} with the two change-aware approaches: \covcap{} and \covcas{} (see \cref{sec:study:indep-vars:ca}).
\Cref{fig:effectiveness:change-awareness} depicts the results of the pair-wise statistical tests displayed per project and \gls{smbp} technique.
For each project and technique, the figure shows which change-awareness approaches are statistically better than which other approaches.

\paragraph{Overall}
Contrary to intuition, i.e., a change-aware approach performs better, change-awareness \emph{does not} improve \gls{smbp} effectiveness (see the top row of \cref{fig:effectiveness:change-awareness}). 
To keep the \gls{smbp} techniques simple, i.e., not having to manage code change information, this result suggests that \covnca{} is the best choice overall.

\paragraph{Per Project}
From the overall results, we observe differences among the change-aware approaches, although with no best approach across \gls{smbp} techniques and projects.
Specifically, greedy techniques are less sensitive to change-awareness, i.e., change-awareness has hardly an impact on the effectiveness.
Whereas the \gls{sbsmbp} techniques are considerably impacted by the change-awareness choice, each technique is not equally affected by the same approach.
We notice that change-awareness is connected to the project that the \gls{smbp} is applied to, e.g., using \covcap{} for \bytebuddy{}, \covcas{} for \jctools{}, and \covnca{} for (most \gls{smbp} techniques for) \okio{} is most effective.
This shows that there often is a benefit for practitioners choosing the \enquote{right} change-awareness approach depending on the project and \gls{smbp} technique.

\summarybox{\rqchangeawareness{} Summary}{
    Change-awareness does not impact \gls{smbp} effectiveness overall.
    However, depending on the concrete project and \gls{smbp} technique, one of the three change-awareness approaches can be more effective, in particular when using a \gls{sbsmbp} technique.
}
  
\subsection{\rqefficiency{}: Efficiency}
\label{sec:results:efficiency}

While the previous section investigated the effectiveness of the \gls{sbsmbp} and \saGreedy{} techniques, a holistic evaluation requires an efficiency evaluation to understand whether using a technique is feasible in practice.
This section compares the greedy heuristics \prioadditional{} and \priototal{}, \saGreedy{} techniques, and \gls{sbsmbp} techniques with respect to the overhead they impose on the overall benchmark suite execution time.
\Cref{tab:efficiency} shows the median and \gls{mad} efficiency overheads overall and per project.
Overheads below \num{1}\% are depicted as \enquote{$<1\%$}.
We refrain from reporting statistical tests, as they would arguably neither add valuable insights nor change the conclusions.

\begin{table*}[tbp]
    \centering
    \tiny

    \addtolength{\extrarowheight}{\belowrulesep}
    \aboverulesep=0pt
    \belowrulesep=0pt

    \caption{Runtime overhead (prioritization and analysis time) for all the projects and overall for the \gls{smbp} techniques}
    \label{tab:efficiency}

    \begin{tabular}{lllllllllllllll}
\toprule
\gls{smbp} Algorithm & Objectives & Time & Overall & \multicolumn{10}{l}{Projects} \\
\cmidrule(l{1pt}r{1pt}){5-14}
 & & & & \bytebuddy{} & \eclipsecollections{} & \jctools{} & \jenetics{} & \logging{} & \netty{} & \okio{} & \rxjava{} & \xodus{} & \zipkin{} \\
\midrule

\rowcolor{lightgray!20}
 \priototal{} & -- & Prioritization & \textless{}\num{1}\%$\pm$\textless{}\num{1} & \textless{}\num{1}\%$\pm$\textless{}\num{1} & \textless{}\num{1}\%$\pm$\textless{}\num{1} & \textless{}\num{1}\%$\pm$\textless{}\num{1} & \textless{}\num{1}\%$\pm$\textless{}\num{1} & \textless{}\num{1}\%$\pm$\textless{}\num{1} & \textless{}\num{1}\%$\pm$\textless{}\num{1} & \textless{}\num{1}\%$\pm$\textless{}\num{1} & \textless{}\num{1}\%$\pm$\textless{}\num{1} & \textless{}\num{1}\%$\pm$\textless{}\num{1} & \textless{}\num{1}\%$\pm$\textless{}\num{1} \\
\rowcolor{lightgray!20}
  &  & Analysis & \num{17}\%$\pm$\textless{}\num{1} & \num{13}\%$\pm$\textless{}\num{1} & \num{104}\%$\pm$\textless{}\num{1} & \num{19}\%$\pm$\textless{}\num{1} & \num{8}\%$\pm$\textless{}\num{1} & \num{22}\%$\pm$\textless{}\num{1} & \num{49}\%$\pm$\textless{}\num{1} & \num{14}\%$\pm$\textless{}\num{1} & \num{23}\%$\pm$\textless{}\num{1} & \num{16}\%$\pm$\textless{}\num{1} & \num{20}\%$\pm$\textless{}\num{1} \\
\rowcolor{white}
 \prioadditional{} & -- & Prioritization & \textless{}\num{1}\%$\pm$\textless{}\num{1} & \textless{}\num{1}\%$\pm$\textless{}\num{1} & \textless{}\num{1}\%$\pm$\textless{}\num{1} & \textless{}\num{1}\%$\pm$\textless{}\num{1} & \textless{}\num{1}\%$\pm$\textless{}\num{1} & \textless{}\num{1}\%$\pm$\textless{}\num{1} & \num{1}\%$\pm$\num{1} & \textless{}\num{1}\%$\pm$\textless{}\num{1} & \textless{}\num{1}\%$\pm$\textless{}\num{1} & \textless{}\num{1}\%$\pm$\textless{}\num{1} & \textless{}\num{1}\%$\pm$\textless{}\num{1} \\
\rowcolor{white}
  &  & Analysis & \num{17}\%$\pm$\textless{}\num{1} & \num{13}\%$\pm$\textless{}\num{1} & \num{105}\%$\pm$\textless{}\num{1} & \num{19}\%$\pm$\textless{}\num{1} & \num{8}\%$\pm$\textless{}\num{1} & \num{22}\%$\pm$\textless{}\num{1} & \num{50}\%$\pm$\num{1} & \num{14}\%$\pm$\textless{}\num{1} & \num{23}\%$\pm$\textless{}\num{1} & \num{16}\%$\pm$\textless{}\num{1} & \num{20}\%$\pm$\textless{}\num{1} \\
\rowcolor{lightgray!20}
 \saGreedy{} & \objsc{} & Prioritization & \textless{}\num{1}\%$\pm$\textless{}\num{1} & \textless{}\num{1}\%$\pm$\textless{}\num{1} & \textless{}\num{1}\%$\pm$\textless{}\num{1} & \textless{}\num{1}\%$\pm$\textless{}\num{1} & \textless{}\num{1}\%$\pm$\textless{}\num{1} & \textless{}\num{1}\%$\pm$\textless{}\num{1} & \textless{}\num{1}\%$\pm$\textless{}\num{1} & \textless{}\num{1}\%$\pm$\textless{}\num{1} & \textless{}\num{1}\%$\pm$\textless{}\num{1} & \textless{}\num{1}\%$\pm$\textless{}\num{1} & \textless{}\num{1}\%$\pm$\textless{}\num{1} \\
\rowcolor{lightgray!20}
  &  & Analysis & \num{17}\%$\pm$\textless{}\num{1} & \num{13}\%$\pm$\textless{}\num{1} & \num{104}\%$\pm$\textless{}\num{1} & \num{19}\%$\pm$\textless{}\num{1} & \num{8}\%$\pm$\textless{}\num{1} & \num{22}\%$\pm$\textless{}\num{1} & \num{49}\%$\pm$\textless{}\num{1} & \num{14}\%$\pm$\textless{}\num{1} & \num{23}\%$\pm$\textless{}\num{1} & \num{16}\%$\pm$\textless{}\num{1} & \num{20}\%$\pm$\textless{}\num{1} \\
\rowcolor{white}
  & \objsco{} & Prioritization & \textless{}\num{1}\%$\pm$\textless{}\num{1} & \textless{}\num{1}\%$\pm$\textless{}\num{1} & \textless{}\num{1}\%$\pm$\textless{}\num{1} & \textless{}\num{1}\%$\pm$\textless{}\num{1} & \textless{}\num{1}\%$\pm$\textless{}\num{1} & \textless{}\num{1}\%$\pm$\textless{}\num{1} & \textless{}\num{1}\%$\pm$\textless{}\num{1} & \textless{}\num{1}\%$\pm$\textless{}\num{1} & \textless{}\num{1}\%$\pm$\textless{}\num{1} & \textless{}\num{1}\%$\pm$\textless{}\num{1} & \textless{}\num{1}\%$\pm$\textless{}\num{1} \\
\rowcolor{white}
  &  & Analysis & \num{17}\%$\pm$\textless{}\num{1} & \num{13}\%$\pm$\textless{}\num{1} & \num{104}\%$\pm$\textless{}\num{1} & \num{19}\%$\pm$\textless{}\num{1} & \num{8}\%$\pm$\textless{}\num{1} & \num{22}\%$\pm$\textless{}\num{1} & \num{49}\%$\pm$\textless{}\num{1} & \num{14}\%$\pm$\textless{}\num{1} & \num{23}\%$\pm$\textless{}\num{1} & \num{16}\%$\pm$\textless{}\num{1} & \num{20}\%$\pm$\textless{}\num{1} \\
\rowcolor{lightgray!20}
  & \objsch{} & Prioritization & \textless{}\num{1}\%$\pm$\textless{}\num{1} & \textless{}\num{1}\%$\pm$\textless{}\num{1} & \textless{}\num{1}\%$\pm$\textless{}\num{1} & \textless{}\num{1}\%$\pm$\textless{}\num{1} & \textless{}\num{1}\%$\pm$\textless{}\num{1} & \textless{}\num{1}\%$\pm$\textless{}\num{1} & \textless{}\num{1}\%$\pm$\textless{}\num{1} & \textless{}\num{1}\%$\pm$\textless{}\num{1} & \textless{}\num{1}\%$\pm$\textless{}\num{1} & \textless{}\num{1}\%$\pm$\textless{}\num{1} & \textless{}\num{1}\%$\pm$\textless{}\num{1} \\
\rowcolor{lightgray!20}
  &  & Analysis & \textless{}\num{1}\%$\pm$\textless{}\num{1} & \textless{}\num{1}\%$\pm$\textless{}\num{1} & \textless{}\num{1}\%$\pm$\textless{}\num{1} & \textless{}\num{1}\%$\pm$\textless{}\num{1} & \textless{}\num{1}\%$\pm$\textless{}\num{1} & \textless{}\num{1}\%$\pm$\textless{}\num{1} & \textless{}\num{1}\%$\pm$\textless{}\num{1} & \textless{}\num{1}\%$\pm$\textless{}\num{1} & \textless{}\num{1}\%$\pm$\textless{}\num{1} & \textless{}\num{1}\%$\pm$\textless{}\num{1} & \textless{}\num{1}\%$\pm$\textless{}\num{1} \\
\rowcolor{white}
  & \objsccoch{} & Prioritization & \textless{}\num{1}\%$\pm$\textless{}\num{1} & \textless{}\num{1}\%$\pm$\textless{}\num{1} & \textless{}\num{1}\%$\pm$\textless{}\num{1} & \textless{}\num{1}\%$\pm$\textless{}\num{1} & \textless{}\num{1}\%$\pm$\textless{}\num{1} & \textless{}\num{1}\%$\pm$\textless{}\num{1} & \textless{}\num{1}\%$\pm$\textless{}\num{1} & \textless{}\num{1}\%$\pm$\textless{}\num{1} & \textless{}\num{1}\%$\pm$\textless{}\num{1} & \textless{}\num{1}\%$\pm$\textless{}\num{1} & \textless{}\num{1}\%$\pm$\textless{}\num{1} \\
\rowcolor{white}
  &  & Analysis & \num{17}\%$\pm$\textless{}\num{1} & \num{13}\%$\pm$\textless{}\num{1} & \num{104}\%$\pm$\textless{}\num{1} & \num{19}\%$\pm$\textless{}\num{1} & \num{8}\%$\pm$\textless{}\num{1} & \num{22}\%$\pm$\textless{}\num{1} & \num{49}\%$\pm$\textless{}\num{1} & \num{14}\%$\pm$\textless{}\num{1} & \num{23}\%$\pm$\textless{}\num{1} & \num{16}\%$\pm$\textless{}\num{1} & \num{20}\%$\pm$\textless{}\num{1} \\
\rowcolor{lightgray!20}
 \saHC{} & \objsc{} & Prioritization & \textless{}\num{1}\%$\pm$\textless{}\num{1} & \textless{}\num{1}\%$\pm$\textless{}\num{1} & \textless{}\num{1}\%$\pm$\textless{}\num{1} & \textless{}\num{1}\%$\pm$\textless{}\num{1} & \textless{}\num{1}\%$\pm$\textless{}\num{1} & \textless{}\num{1}\%$\pm$\textless{}\num{1} & \textless{}\num{1}\%$\pm$\textless{}\num{1} & \textless{}\num{1}\%$\pm$\textless{}\num{1} & \textless{}\num{1}\%$\pm$\textless{}\num{1} & \textless{}\num{1}\%$\pm$\textless{}\num{1} & \textless{}\num{1}\%$\pm$\textless{}\num{1} \\
\rowcolor{lightgray!20}
  &  & Analysis & \num{17}\%$\pm$\textless{}\num{1} & \num{13}\%$\pm$\textless{}\num{1} & \num{104}\%$\pm$\textless{}\num{1} & \num{19}\%$\pm$\textless{}\num{1} & \num{8}\%$\pm$\textless{}\num{1} & \num{22}\%$\pm$\textless{}\num{1} & \num{49}\%$\pm$\textless{}\num{1} & \num{14}\%$\pm$\textless{}\num{1} & \num{23}\%$\pm$\textless{}\num{1} & \num{16}\%$\pm$\textless{}\num{1} & \num{20}\%$\pm$\textless{}\num{1} \\
\rowcolor{white}
  & \objsco{} & Prioritization & \textless{}\num{1}\%$\pm$\textless{}\num{1} & \textless{}\num{1}\%$\pm$\textless{}\num{1} & \textless{}\num{1}\%$\pm$\textless{}\num{1} & \textless{}\num{1}\%$\pm$\textless{}\num{1} & \textless{}\num{1}\%$\pm$\textless{}\num{1} & \textless{}\num{1}\%$\pm$\textless{}\num{1} & \textless{}\num{1}\%$\pm$\textless{}\num{1} & \textless{}\num{1}\%$\pm$\textless{}\num{1} & \textless{}\num{1}\%$\pm$\textless{}\num{1} & \textless{}\num{1}\%$\pm$\textless{}\num{1} & \textless{}\num{1}\%$\pm$\textless{}\num{1} \\
\rowcolor{white}
  &  & Analysis & \num{17}\%$\pm$\textless{}\num{1} & \num{13}\%$\pm$\textless{}\num{1} & \num{104}\%$\pm$\textless{}\num{1} & \num{19}\%$\pm$\textless{}\num{1} & \num{8}\%$\pm$\textless{}\num{1} & \num{22}\%$\pm$\textless{}\num{1} & \num{49}\%$\pm$\textless{}\num{1} & \num{14}\%$\pm$\textless{}\num{1} & \num{23}\%$\pm$\textless{}\num{1} & \num{16}\%$\pm$\textless{}\num{1} & \num{20}\%$\pm$\textless{}\num{1} \\
\rowcolor{lightgray!20}
  & \objsch{} & Prioritization & \textless{}\num{1}\%$\pm$\textless{}\num{1} & \textless{}\num{1}\%$\pm$\textless{}\num{1} & \textless{}\num{1}\%$\pm$\textless{}\num{1} & \textless{}\num{1}\%$\pm$\textless{}\num{1} & \textless{}\num{1}\%$\pm$\textless{}\num{1} & \textless{}\num{1}\%$\pm$\textless{}\num{1} & \textless{}\num{1}\%$\pm$\textless{}\num{1} & \textless{}\num{1}\%$\pm$\textless{}\num{1} & \textless{}\num{1}\%$\pm$\textless{}\num{1} & \textless{}\num{1}\%$\pm$\textless{}\num{1} & \textless{}\num{1}\%$\pm$\textless{}\num{1} \\
\rowcolor{lightgray!20}
  &  & Analysis & \textless{}\num{1}\%$\pm$\textless{}\num{1} & \textless{}\num{1}\%$\pm$\textless{}\num{1} & \textless{}\num{1}\%$\pm$\textless{}\num{1} & \textless{}\num{1}\%$\pm$\textless{}\num{1} & \textless{}\num{1}\%$\pm$\textless{}\num{1} & \textless{}\num{1}\%$\pm$\textless{}\num{1} & \textless{}\num{1}\%$\pm$\textless{}\num{1} & \textless{}\num{1}\%$\pm$\textless{}\num{1} & \textless{}\num{1}\%$\pm$\textless{}\num{1} & \textless{}\num{1}\%$\pm$\textless{}\num{1} & \textless{}\num{1}\%$\pm$\textless{}\num{1} \\
\rowcolor{white}
  & \objsccoch{} & Prioritization & \textless{}\num{1}\%$\pm$\textless{}\num{1} & \textless{}\num{1}\%$\pm$\textless{}\num{1} & \textless{}\num{1}\%$\pm$\textless{}\num{1} & \textless{}\num{1}\%$\pm$\textless{}\num{1} & \textless{}\num{1}\%$\pm$\textless{}\num{1} & \textless{}\num{1}\%$\pm$\textless{}\num{1} & \textless{}\num{1}\%$\pm$\textless{}\num{1} & \textless{}\num{1}\%$\pm$\textless{}\num{1} & \textless{}\num{1}\%$\pm$\textless{}\num{1} & \textless{}\num{1}\%$\pm$\textless{}\num{1} & \textless{}\num{1}\%$\pm$\textless{}\num{1} \\
\rowcolor{white}
  &  & Analysis & \num{17}\%$\pm$\textless{}\num{1} & \num{13}\%$\pm$\textless{}\num{1} & \num{104}\%$\pm$\textless{}\num{1} & \num{19}\%$\pm$\textless{}\num{1} & \num{8}\%$\pm$\textless{}\num{1} & \num{22}\%$\pm$\textless{}\num{1} & \num{49}\%$\pm$\textless{}\num{1} & \num{14}\%$\pm$\textless{}\num{1} & \num{23}\%$\pm$\textless{}\num{1} & \num{16}\%$\pm$\textless{}\num{1} & \num{20}\%$\pm$\textless{}\num{1} \\
\rowcolor{lightgray!20}
 \saGA{} & \objsc{} & Prioritization & \num{1}\%$\pm$\num{1} & \num{2}\%$\pm$\num{1} & \textless{}\num{1}\%$\pm$\textless{}\num{1} & \textless{}\num{1}\%$\pm$\textless{}\num{1} & \num{1}\%$\pm$\textless{}\num{1} & \textless{}\num{1}\%$\pm$\textless{}\num{1} & \textless{}\num{1}\%$\pm$\textless{}\num{1} & \textless{}\num{1}\%$\pm$\textless{}\num{1} & \textless{}\num{1}\%$\pm$\textless{}\num{1} & \textless{}\num{1}\%$\pm$\textless{}\num{1} & \num{1}\%$\pm$\textless{}\num{1} \\
\rowcolor{lightgray!20}
  &  & Analysis & \num{18}\%$\pm$\num{1} & \num{15}\%$\pm$\num{1} & \num{104}\%$\pm$\textless{}\num{1} & \num{19}\%$\pm$\textless{}\num{1} & \num{9}\%$\pm$\textless{}\num{1} & \num{23}\%$\pm$\textless{}\num{1} & \num{49}\%$\pm$\textless{}\num{1} & \num{14}\%$\pm$\textless{}\num{1} & \num{23}\%$\pm$\textless{}\num{1} & \num{16}\%$\pm$\textless{}\num{1} & \num{22}\%$\pm$\textless{}\num{1} \\
\rowcolor{white}
  & \objsco{} & Prioritization & \num{1}\%$\pm$\num{1} & \num{2}\%$\pm$\num{1} & \textless{}\num{1}\%$\pm$\textless{}\num{1} & \textless{}\num{1}\%$\pm$\textless{}\num{1} & \num{1}\%$\pm$\textless{}\num{1} & \textless{}\num{1}\%$\pm$\textless{}\num{1} & \textless{}\num{1}\%$\pm$\textless{}\num{1} & \textless{}\num{1}\%$\pm$\textless{}\num{1} & \textless{}\num{1}\%$\pm$\textless{}\num{1} & \textless{}\num{1}\%$\pm$\textless{}\num{1} & \num{1}\%$\pm$\textless{}\num{1} \\
\rowcolor{white}
  &  & Analysis & \num{19}\%$\pm$\num{1} & \num{15}\%$\pm$\num{1} & \num{104}\%$\pm$\textless{}\num{1} & \num{19}\%$\pm$\textless{}\num{1} & \num{9}\%$\pm$\textless{}\num{1} & \num{23}\%$\pm$\textless{}\num{1} & \num{49}\%$\pm$\textless{}\num{1} & \num{14}\%$\pm$\textless{}\num{1} & \num{23}\%$\pm$\textless{}\num{1} & \num{16}\%$\pm$\textless{}\num{1} & \num{22}\%$\pm$\textless{}\num{1} \\
\rowcolor{lightgray!20}
  & \objsch{} & Prioritization & \num{1}\%$\pm$\num{1} & \num{2}\%$\pm$\num{1} & \textless{}\num{1}\%$\pm$\textless{}\num{1} & \textless{}\num{1}\%$\pm$\textless{}\num{1} & \num{1}\%$\pm$\textless{}\num{1} & \textless{}\num{1}\%$\pm$\textless{}\num{1} & \textless{}\num{1}\%$\pm$\textless{}\num{1} & \textless{}\num{1}\%$\pm$\textless{}\num{1} & \textless{}\num{1}\%$\pm$\textless{}\num{1} & \textless{}\num{1}\%$\pm$\textless{}\num{1} & \num{1}\%$\pm$\textless{}\num{1} \\
\rowcolor{lightgray!20}
  &  & Analysis & \num{1}\%$\pm$\num{1} & \num{2}\%$\pm$\num{1} & \textless{}\num{1}\%$\pm$\textless{}\num{1} & \textless{}\num{1}\%$\pm$\textless{}\num{1} & \num{1}\%$\pm$\textless{}\num{1} & \textless{}\num{1}\%$\pm$\textless{}\num{1} & \textless{}\num{1}\%$\pm$\textless{}\num{1} & \textless{}\num{1}\%$\pm$\textless{}\num{1} & \textless{}\num{1}\%$\pm$\textless{}\num{1} & \textless{}\num{1}\%$\pm$\textless{}\num{1} & \num{1}\%$\pm$\textless{}\num{1} \\
\rowcolor{white}
  & \objsccoch{} & Prioritization & \num{1}\%$\pm$\num{1} & \num{2}\%$\pm$\num{1} & \textless{}\num{1}\%$\pm$\textless{}\num{1} & \num{1}\%$\pm$\textless{}\num{1} & \num{2}\%$\pm$\textless{}\num{1} & \textless{}\num{1}\%$\pm$\textless{}\num{1} & \textless{}\num{1}\%$\pm$\textless{}\num{1} & \textless{}\num{1}\%$\pm$\textless{}\num{1} & \textless{}\num{1}\%$\pm$\textless{}\num{1} & \textless{}\num{1}\%$\pm$\textless{}\num{1} & \num{1}\%$\pm$\textless{}\num{1} \\
\rowcolor{white}
  &  & Analysis & \num{19}\%$\pm$\num{1} & \num{15}\%$\pm$\num{1} & \num{104}\%$\pm$\textless{}\num{1} & \num{20}\%$\pm$\textless{}\num{1} & \num{9}\%$\pm$\textless{}\num{1} & \num{23}\%$\pm$\textless{}\num{1} & \num{49}\%$\pm$\textless{}\num{1} & \num{14}\%$\pm$\textless{}\num{1} & \num{23}\%$\pm$\textless{}\num{1} & \num{16}\%$\pm$\textless{}\num{1} & \num{23}\%$\pm$\textless{}\num{1} \\
\rowcolor{lightgray!20}
 \saIBEA{} & \objsccoch{} & Prioritization & \num{1}\%$\pm$\num{2} & \num{5}\%$\pm$\num{2} & \textless{}\num{1}\%$\pm$\textless{}\num{1} & \num{1}\%$\pm$\textless{}\num{1} & \num{3}\%$\pm$\textless{}\num{1} & \num{1}\%$\pm$\textless{}\num{1} & \textless{}\num{1}\%$\pm$\textless{}\num{1} & \num{1}\%$\pm$\textless{}\num{1} & \textless{}\num{1}\%$\pm$\textless{}\num{1} & \num{1}\%$\pm$\textless{}\num{1} & \num{3}\%$\pm$\num{1} \\
\rowcolor{lightgray!20}
  &  & Analysis & \num{20}\%$\pm$\num{2} & \num{18}\%$\pm$\num{2} & \num{104}\%$\pm$\textless{}\num{1} & \num{20}\%$\pm$\textless{}\num{1} & \num{11}\%$\pm$\textless{}\num{1} & \num{23}\%$\pm$\textless{}\num{1} & \num{49}\%$\pm$\textless{}\num{1} & \num{15}\%$\pm$\textless{}\num{1} & \num{23}\%$\pm$\textless{}\num{1} & \num{17}\%$\pm$\textless{}\num{1} & \num{25}\%$\pm$\num{1} \\
\rowcolor{white}
 \saMOCell{} & \objsccoch{} & Prioritization & \textless{}\num{1}\%$\pm$\textless{}\num{1} & \num{3}\%$\pm$\num{1} & \textless{}\num{1}\%$\pm$\textless{}\num{1} & \textless{}\num{1}\%$\pm$\textless{}\num{1} & \num{1}\%$\pm$\textless{}\num{1} & \textless{}\num{1}\%$\pm$\textless{}\num{1} & \textless{}\num{1}\%$\pm$\textless{}\num{1} & \textless{}\num{1}\%$\pm$\textless{}\num{1} & \textless{}\num{1}\%$\pm$\textless{}\num{1} & \textless{}\num{1}\%$\pm$\textless{}\num{1} & \num{1}\%$\pm$\textless{}\num{1} \\
\rowcolor{white}
  &  & Analysis & \num{18}\%$\pm$\textless{}\num{1} & \num{15}\%$\pm$\num{1} & \num{105}\%$\pm$\textless{}\num{1} & \num{19}\%$\pm$\textless{}\num{1} & \num{9}\%$\pm$\textless{}\num{1} & \num{22}\%$\pm$\textless{}\num{1} & \num{49}\%$\pm$\textless{}\num{1} & \num{14}\%$\pm$\textless{}\num{1} & \num{23}\%$\pm$\textless{}\num{1} & \num{16}\%$\pm$\textless{}\num{1} & \num{23}\%$\pm$\textless{}\num{1} \\
\rowcolor{lightgray!20}
 \saNSGAII{} & \objsccoch{} & Prioritization & \num{1}\%$\pm$\num{1} & \num{2}\%$\pm$\num{1} & \textless{}\num{1}\%$\pm$\textless{}\num{1} & \textless{}\num{1}\%$\pm$\textless{}\num{1} & \num{1}\%$\pm$\textless{}\num{1} & \textless{}\num{1}\%$\pm$\textless{}\num{1} & \textless{}\num{1}\%$\pm$\textless{}\num{1} & \textless{}\num{1}\%$\pm$\textless{}\num{1} & \textless{}\num{1}\%$\pm$\textless{}\num{1} & \textless{}\num{1}\%$\pm$\textless{}\num{1} & \num{1}\%$\pm$\textless{}\num{1} \\
\rowcolor{lightgray!20}
  &  & Analysis & \num{19}\%$\pm$\num{1} & \num{15}\%$\pm$\num{1} & \num{104}\%$\pm$\textless{}\num{1} & \num{20}\%$\pm$\textless{}\num{1} & \num{9}\%$\pm$\textless{}\num{1} & \num{23}\%$\pm$\textless{}\num{1} & \num{49}\%$\pm$\textless{}\num{1} & \num{14}\%$\pm$\textless{}\num{1} & \num{23}\%$\pm$\textless{}\num{1} & \num{16}\%$\pm$\textless{}\num{1} & \num{22}\%$\pm$\textless{}\num{1} \\
\rowcolor{white}
 \saNSGAIII{} & \objsccoch{} & Prioritization & \textless{}\num{1}\%$\pm$\textless{}\num{1} & \textless{}\num{1}\%$\pm$\textless{}\num{1} & \textless{}\num{1}\%$\pm$\textless{}\num{1} & \textless{}\num{1}\%$\pm$\textless{}\num{1} & \textless{}\num{1}\%$\pm$\textless{}\num{1} & \textless{}\num{1}\%$\pm$\textless{}\num{1} & \textless{}\num{1}\%$\pm$\textless{}\num{1} & \textless{}\num{1}\%$\pm$\textless{}\num{1} & \textless{}\num{1}\%$\pm$\textless{}\num{1} & \textless{}\num{1}\%$\pm$\textless{}\num{1} & \textless{}\num{1}\%$\pm$\textless{}\num{1} \\
\rowcolor{white}
  &  & Analysis & \num{17}\%$\pm$\textless{}\num{1} & \num{13}\%$\pm$\textless{}\num{1} & \num{104}\%$\pm$\textless{}\num{1} & \num{19}\%$\pm$\textless{}\num{1} & \num{8}\%$\pm$\textless{}\num{1} & \num{22}\%$\pm$\textless{}\num{1} & \num{49}\%$\pm$\textless{}\num{1} & \num{14}\%$\pm$\textless{}\num{1} & \num{23}\%$\pm$\textless{}\num{1} & \num{16}\%$\pm$\textless{}\num{1} & \num{21}\%$\pm$\textless{}\num{1} \\
\rowcolor{lightgray!20}
 \saPAES{} & \objsccoch{} & Prioritization & \textless{}\num{1}\%$\pm$\textless{}\num{1} & \textless{}\num{1}\%$\pm$\textless{}\num{1} & \textless{}\num{1}\%$\pm$\textless{}\num{1} & \textless{}\num{1}\%$\pm$\textless{}\num{1} & \textless{}\num{1}\%$\pm$\textless{}\num{1} & \textless{}\num{1}\%$\pm$\textless{}\num{1} & \textless{}\num{1}\%$\pm$\textless{}\num{1} & \textless{}\num{1}\%$\pm$\textless{}\num{1} & \textless{}\num{1}\%$\pm$\textless{}\num{1} & \textless{}\num{1}\%$\pm$\textless{}\num{1} & \textless{}\num{1}\%$\pm$\textless{}\num{1} \\
\rowcolor{lightgray!20}
  &  & Analysis & \num{17}\%$\pm$\textless{}\num{1} & \num{13}\%$\pm$\textless{}\num{1} & \num{104}\%$\pm$\textless{}\num{1} & \num{19}\%$\pm$\textless{}\num{1} & \num{8}\%$\pm$\textless{}\num{1} & \num{22}\%$\pm$\textless{}\num{1} & \num{49}\%$\pm$\textless{}\num{1} & \num{14}\%$\pm$\textless{}\num{1} & \num{23}\%$\pm$\textless{}\num{1} & \num{16}\%$\pm$\textless{}\num{1} & \num{21}\%$\pm$\textless{}\num{1} \\
\rowcolor{white}
 \saSPEA{} & \objsccoch{} & Prioritization & \num{2}\%$\pm$\num{2} & \num{9}\%$\pm$\num{3} & \textless{}\num{1}\%$\pm$\textless{}\num{1} & \num{1}\%$\pm$\textless{}\num{1} & \num{5}\%$\pm$\num{1} & \num{1}\%$\pm$\textless{}\num{1} & \textless{}\num{1}\%$\pm$\textless{}\num{1} & \num{1}\%$\pm$\textless{}\num{1} & \textless{}\num{1}\%$\pm$\textless{}\num{1} & \num{1}\%$\pm$\textless{}\num{1} & \num{3}\%$\pm$\num{1} \\
\rowcolor{white}
  &  & Analysis & \num{22}\%$\pm$\num{2} & \num{22}\%$\pm$\num{3} & \num{105}\%$\pm$\textless{}\num{1} & \num{20}\%$\pm$\textless{}\num{1} & \num{13}\%$\pm$\num{1} & \num{23}\%$\pm$\textless{}\num{1} & \num{49}\%$\pm$\textless{}\num{1} & \num{15}\%$\pm$\textless{}\num{1} & \num{23}\%$\pm$\textless{}\num{1} & \num{17}\%$\pm$\textless{}\num{1} & \num{26}\%$\pm$\num{1} \\

\bottomrule
\end{tabular}
 \end{table*}
 
\paragraph{Overall}
Both baselines add less than \num{1}\% for running the algorithm, i.e., prioritization time, and combined with the coverage extraction time have a total runtime overhead, i.e., analysis time, of \num{17}\%.
This shows that the analysis time is dominated by the coverage extraction and the \gls{smbp} algorithm plays only a minor role.
These numbers are from our previous work~\citep{laaber:21c}.
The same applies to the coverage-based \saGreedy{} techniques, i.e., \objsc{}, \objsco{}, and \objsccoch{}.

The single-objective \gls{sbsmbp} techniques (i.e., \gls{hc} and \gls{ga}) have around \num{1}\% prioritization time.
This brings the analysis time to between \num{17}\% and \num{19}\%, similar to the greedy baselines.
\saGreedy{} \objsch{}, \gls{hc} \objsch{}, and \gls{ga} \objsch{}, however, have a similar analysis and prioritization time.
This is because the \objsch{} objective does not rely on coverage extraction, which takes the majority of the analysis time for coverage-based techniques.
In particular, \saGreedy{} \objsch{} becomes (even more) appealing, as it is not only among the most effective techniques (with, e.g., \priototal{}) but also the most efficient one.

Regarding the multi-objective \gls{sbsmbp} technique, we observe that the overall overheads do not change for the majority of the \glspl{moea}, i.e., \gls{mocell} (the most effective \gls{moea} from \rqeffectiveness{}, see \cref{sec:results:effectiveness}), \gls{nsgaii}, \gls{nsgaiii}, and \gls{paes}.
Only \gls{ibea} and \gls{spea2} encounter a slightly larger prioritization time,
which increases the analysis time to \num{19}\% and \num{21}\%, respectively. Given the large suite runtimes (see \cref{tab:study-objects}), this slight increase over the greedy techniques can be considered acceptable.

\paragraph{Per Project}
The results per project show a diverse situation for the \gls{smbp} techniques that rely on coverage objectives: the analysis times range from between \num{8}\% and \num{13}\% for \jenetics{} to between \num{104}\% and \num{105}\% for \eclipsecollections{}.
For eight of the projects, the analysis time is \num{23}\% or less.
For \eclipsecollections{} and \netty{}, where the overhead is \num{105}\% and \num{49}\% respectively, applying \gls{smbp} is not worthwhile.

These project-dependent overheads suggest that the techniques with non-coverage-based objectives are critical to be universally applicable across projects, especially when run as part of \gls{ci}.
Hence, the overhead results for the \objsch{} techniques are especially promising:
all \objsch{} techniques have an analysis time overhead of \num{2}\% or less across all the projects.
When considering the most effective technique \saGreedy{} \objsch{}, the overhead even drops below \num{1}\% for all the projects.
Consequently, the analysis time of the techniques with non-coverage-based objectives is solely dependent on the prioritization time (i.e., the time taken to run the \gls{smbp} algorithm), which will likely always be considerably smaller than the extensive benchmark suite runtimes; therefore, it is generally beneficial to apply these \gls{smbp} techniques.

\summarybox{\rqefficiency{} Summary}{
    The \gls{sbsmbp} techniques impose only a minor additional overhead compared to the greedy baselines.
    Employing a \gls{smbp} technique that only relies on the \objsch{} objective gives the lowest and most reliable overhead across all the projects.
    In particular, \saGreedy{} \objsch{} consistently has less than \num{1}\% overhead.
}
  
\section{Discussion}
\label{sec:discussion}

The results show that the best \gls{sbsmbp} technique (\gls{ga} \objsccoch{}) is only competitive with the best greedy baseline (\priototal{}).
When using a simple \saGreedy{} technique that only considers the historical performance change size (\objsch{}), one can retrieve equally and sometimes more effective \gls{smbp} rankings with significantly lower overhead than when relying on coverage objectives (\objsc{} and \objsco{}), which is the case for both the greedy baselines and the \gls{sbsmbp} techniques.
This section discusses our empirical results from several different perspectives.

\paragraph{The Underperforming Search-Based Techniques}
The success of the \saGreedy{} \objsch{} technique in terms of effectiveness begs the question, why are the \gls{sbsmbp} techniques underperforming in our study?
We see the following two reasons.
    
First, coverage is not \enquote{good enough} as objectives for \gls{smbp}.
This becomes evident from this and our previous study~\citep{laaber:21c}.
Furthermore, \citet{chen:22} observed that in the context of benchmark selection, functional bug prediction metrics are more important than code-level performance metrics.
This suggests that future \gls{smbp} techniques should explore dedicated metrics related to size, diffusion, or history as well as specific code-level performance metrics, such as loops or synchronization (similar to \citet{laaber:21b}).
However, given the high coverage extraction overhead, non-code metrics should be more favoured over code metrics.

Second, optimal hyperparameters of the \glspl{sa} could improve \gls{sbsmbp} effectiveness.
For this, we explored the \glspl{sa}' effectiveness with more generations, investigated the convergence of the objectives, and found that the objectives converge quickly.
This suggests that the search space is relatively trivial (for the given objectives) and \glspl{sa} do not offer an advantage over simple greedy techniques, which our results empirically show.
Moreover, the employed search objectives, in particular the coverage-based objectives, are ineffective in finding larger performance changes sooner, and other objectives should be explored in the future (such as the ones mentioned above).

\paragraphquestion{Is it Worth Applying \gls{smbp}}
While coverage-based \gls{smbp} is effective, there is still a considerable overhead.
This overhead is mostly due to the time required to extract coverage information~\citep{laaber:21c} and only a small fraction is attributed to the prioritization algorithm (see \cref{sec:results:efficiency}).
While this time is arguably lower than for unit testing because benchmarks are repeatedly executed and run much longer~\citep{laaber:20}, whether it is worth employing coverage-based \gls{smbp} highly depends on the individual benchmark suite and the project's performance testing objectives.
If it is critical to detect performance changes as fast as possible, e.g., the release of a new software version would be otherwise halted, running \gls{smbp} can drastically reduce the time to detect these.
However, for projects with extensive overheads, such as \eclipsecollections{} and \netty{} in our study, coverage-based \gls{smbp} is not worthwhile.
A better alternative to coverage-based techniques are \gls{smbp} techniques that solely rely on the \objsch{} objective, which is (sometimes) more effective and substantially more efficient.
This suggests that non-coverage-based \gls{smbp} techniques are highly suggested to be employed.

\paragraphquestion{Is it Practical to Apply \gls{smbp}}
Beyond the temporal cost of applying a \gls{smbp} technique, there is a cost of extracting and maintaining the information required as input.
On the one hand, the techniques require storing historical information about benchmark executions, especially their changes between previous versions, for computing the \objsch{} objective.
On the other hand, there is neither a need to store historical coverage information nor compute the source code difference on every new commit, as the results of \rqchangeawareness{} show that \covnca{}, the non-change-aware approach, performs equally to the change-aware approaches (i.e., \covcap{} and \covcas{}) overall.
This is probably best done as part of the \gls{ci} pipeline because it provides the necessary infrastructure to run benchmarks, store the information about the build (i.e., performance change history), and has the code changes readily available.

\paragraph{Relying on Measured Performance Changes}
Research on functional \gls{tcp} often uses datasets with seeded, artificial faults (mutations) to evaluate \gls{tcp} effectiveness (e.g., \citep{luo:19,dinucci:20}).
Conversely, performance regression testing research mostly relies on measured performance changes between adjacent versions~\citep{mostafa:17,laaber:21c}.
This is due to performance mutation testing having only recently received attention~\citep{delgado-perez:20,jangali:22} and large datasets, such as Defects4J~\citep{just:14}, do not exist.
Because of the enormous experiment runtimes required for rigorously executing benchmarks (this study's dataset took \num{89} days to create), performing performance mutation testing, where for each mutant the whole benchmark suite has to be executed, becomes unrealistic.
However, this does not invalidate the findings of this study or any other experimental performance regression study relying on measured performance changes as long as the measurements have been conducted rigorously.

Moreover, measuring performance captures changes stemming from \enquote{outside} the \gls{sut}, e.g., its dependencies or runtime environment.
While it is harder for techniques to also detect these changes, it makes the evaluation more realistic, as the \gls{sut}'s developers should be aware of any observable performance changes, irrespective of their origin, to take adequate action.

\paragraphquestion{What is an Important Performance Change}
Following previous research on \gls{smbp}~\citep{mostafa:17,laaber:21c}, this paper also defines the importance of a benchmark as the performance change size it detects.
A benchmark that detects a large performance change is considered more important than a benchmark that detects a small one.
Accordingly, we conclude whether a technique is more or less effective.
It is, however, unclear whether this definition of importance is \enquote{accurate.}
Alternative formulations of importance could be based on:
\begin{inparaenum}
    \item the impact of the code called by the benchmark on the performance of an application,
    \item the project developer's perception and context-dependent knowledge,
    or
    \item the usage frequency of the code from \gls{api} clients.
\end{inparaenum}
Moreover, developers might consider any performance change as important, irrespective of the size.
All these different definitions of importance would likely change our results and require a study of its own.
 
\section{Related Work}
\label{sec:rw}

This section discusses the related work on
\begin{inparaenum}
    \item search-based regression testing
    and
    \item software performance testing.
\end{inparaenum}

\subsection{Search-Based Regression Testing}

Regression testing on unit tests has been a well-established research field for the last \num{30} years~\citep{yoo:12}.
The three main techniques in regression testing are test suite minimization, \gls{rts}, and \gls{tcp}.
Traditionally, techniques relied on greedy heuristics, such as \prioadditional{} and \priototal{} coverage, to solve the optimization problems of selecting the best subset or prioritizing the most fault-revealing tests~\citep{rothermel:98,rothermel:99,elbaum:01}.

Based on the early success of search techniques in software engineering~\citep{harman:01}, regression testing techniques were quickly adapted to use search for the optimization.
\citet{yoo:07} formulate \gls{rts} as a multi-objective optimization problem and generate Pareto optimal solutions~\citep{fonseca:95}.
\citet{li:07} are the first to apply search to \gls{tcp} by introducing coverage-based objectives based on block, decision, and statement granularity.
Our work is inspired by theirs and adapts the objectives for \gls{smbp}.

\citet{li:13} take the idea a step forward and introduce multi-objective, search-based \gls{tcp} with \gls{nsgaii} and two objectives: coverage and execution time.
\citet{islam:12} and \citet{marchetto:16} consider three objectives and apply different weights to these.
\citet{epitropakis:15} address the $\bigo(mn)$ complexity of the fitness function by devising a coverage compaction algorithm and consider historical faults as an objective.
Finally, \citet{dinucci:20} introduce a new search algorithm based on the hypervolume~\citep{auger:09}.
Our work builds on all these in terms of inspiration for the search objectives and the experimental setup (see \cref{sec:study:indep-vars}) and ports them to the problem of prioritizing microbenchmarks.

\subsection{Software Performance Testing}

Performance testing with microbenchmarks is a relatively new area of research.
To this end, earlier studies investigate unit-level performance testing of \gls{oss} Java projects~\citep{leitner:17, stefan:17}, which
conclude that it is still a niche technique, especially compared to unit testing.
Explanations could be that more effort is required to write microbenchmarks, the extra cost of their execution, and the higher complexity required to assess the results with statistical analyses.
\citet{samoaa:21} investigate parameterization of benchmarks in detail.
\citet{grambow:21,grambow:23} utilize information from application benchmarks to guide microbenchmark execution and assess the ability of microbenchmarks to detect application performance changes, with limited success.
\citet{he:19} and \citet{laaber:20} take an orthogonal approach to optimizing performance testing: they dynamically stop benchmarks when their results are of a desired statistical quality instead of optimizing the execution order of the benchmark suite.
\citet{traini:23} further assess whether dynamically stopping is beneficially for measuring in steady state.

There exist works on performance regression testing, i.e., greedy heuristics~\citep{oliveira:17}, genetic algorithms~\citep{alshoaibi:19,alshoaibi:22}, and machine learning~\citep{chen:22}, which have been used to select the optimal benchmarks for a given software version in terms of their ability to find performance changes.
These works focus on \gls{rts} whereas this paper targets \gls{smbp}.
In addition, some works exist for performance regression testing of concurrent classes (e.g., \citep{pradel:14,yu:18}).
They focus on and generate tests for concurrent software, whereas this paper optimizes the execution of existing benchmark suites.
Finally, \citet{huang:14} proposed a prediction method to assess the need for performance testing of a new software version, which targets the version level compared to ours on the benchmark level.

We consider two works on performance regression testing closely related to ours~\citep{mostafa:17,laaber:21c}.
For a given version, \citet{mostafa:17} prioritize test cases with complex performance impact analyses that require measuring individual components of not only the \gls{sut} but also the \gls{jdk}.
Their technique focuses on collection-intensive software and is partially only evaluated on unit tests used as performance tests, whereas our technique is generally applicable to any kind of software and is evaluated only on benchmarks.
\citet{laaber:21c} studied coverage-based, greedy \gls{smbp} techniques and found that they are only marginally more effective than a random ordering.
We build on their work and use \glspl{sa} and novel \saGreedy{} techniques instead.

\section{Conclusions and Future Work}
\label{sec:conclusions}

This paper defines \gls{sbsmbp} and \saGreedy{} \gls{smbp} techniques relying on three objectives: coverage, coverage overlap among benchmarks, and performance change size obtained from historical data.
With an extensive experimental evaluation on \numprojects{} Java projects with \gls{jmh} suites, we study the effectiveness and efficiency of a total of \num{18} new \gls{smbp} techniques and compare these to two coverage-based, greedy \gls{smbp} baselines.
The results reveal that the best \gls{sbsmbp} technique is competitive with the best greedy baselines but does not improve on its effectiveness.
Surprisingly, the \saGreedy{} technique relying only on the historical performance change size is sometimes more but at least equally effective as the best coverage-based, greedy baselines and \gls{sbsmbp} techniques while being significantly more efficient.

This paper provides evidence that \gls{smbp} is a difficult problem to solve and more work is required going forward.
Hence, we envision future research to
\begin{inparaenum}
    \item assess \gls{smbp} on performance mutations and developer-reported performance bugs,
    \item investigate novel algorithms (potentially search-based) specifically targeting \gls{smbp},
    and
    \item devise process- and performance-focused, ideally non-coverage-based, objectives that serve as better proxies for performance changes than the ones studied here.
\end{inparaenum}
 
\section*{Acknowledgements}
The research presented in this paper has received funding from \gls{rcn} under the project \texttt{309642} and has benefited from the \glsentryfull{ex3}, which is supported by the \gls{rcn} project \texttt{270053}.

\footnotesize
\bibliographystyle{IEEEtranSN}
\bibliography{refs.bib,refs-tao-shaukat.bib}

\begin{thebibliography}{86}
\providecommand{\natexlab}[1]{#1}
\providecommand{\url}[1]{#1}
\csname url@samestyle\endcsname
\providecommand{\newblock}{\relax}
\providecommand{\bibinfo}[2]{#2}
\providecommand{\BIBentrySTDinterwordspacing}{\spaceskip=0pt\relax}
\providecommand{\BIBentryALTinterwordstretchfactor}{4}
\providecommand{\BIBentryALTinterwordspacing}{\spaceskip=\fontdimen2\font plus
\BIBentryALTinterwordstretchfactor\fontdimen3\font minus
  \fontdimen4\font\relax}
\providecommand{\BIBforeignlanguage}[2]{{%
\expandafter\ifx\csname l@#1\endcsname\relax
\typeout{** WARNING: IEEEtranSN.bst: No hyphenation pattern has been}%
\typeout{** loaded for the language `#1'. Using the pattern for}%
\typeout{** the default language instead.}%
\else
\language=\csname l@#1\endcsname
\fi
#2}}
\providecommand{\BIBdecl}{\relax}
\BIBdecl

\bibitem[Achimugu et~al.(2014)Achimugu, Selamat, Ibrahim, and
  Mahrin]{achimugu2014systematic}
\BIBentryALTinterwordspacing
P.~Achimugu, A.~Selamat, R.~Ibrahim, and M.~N. Mahrin, ``A systematic
  literature review of software requirements prioritization research,''
  \emph{Information and Software Technology}, vol.~56, no.~6, 2014. [Online].
  Available: \url{https://doi.org/10.1016/j.infsof.2014.02.001}
\BIBentrySTDinterwordspacing

\bibitem[Alshahwan et~al.(2023)Alshahwan, Harman, and Marginean]{alshahwan:23}
\BIBentryALTinterwordspacing
N.~Alshahwan, M.~Harman, and A.~Marginean, ``Software testing research
  challenges: An industrial perspective,'' in \emph{Proceedings of the 16th
  {IEEE} International Conference on Software Testing, Verification and
  Validation}, ser. {ICST} 2023.\hskip 1em plus 0.5em minus 0.4em\relax {IEEE},
  Apr. 2023, pp. 1--10. [Online]. Available:
  \url{https://doi.org/10.1109/ICST57152.2023.00008}
\BIBentrySTDinterwordspacing

\bibitem[Alshoaibi et~al.(2019)Alshoaibi, Hannigan, Gupta, and
  Mkaouer]{alshoaibi:19}
\BIBentryALTinterwordspacing
D.~Alshoaibi, K.~Hannigan, H.~Gupta, and M.~W. Mkaouer, ``{PRICE}: Detection of
  performance regression introducing code changes using static and dynamic
  metrics,'' in \emph{Proceedings of the 11th International Symposium on Search
  Based Software Engineering}, ser. {SSBSE} 2019.\hskip 1em plus 0.5em minus
  0.4em\relax Springer Nature, 2019, pp. 75--88. [Online]. Available:
  \url{https://doi.org/10.1007/978-3-030-27455-9_6}
\BIBentrySTDinterwordspacing

\bibitem[Alshoaibi et~al.(2022)Alshoaibi, Mkaouer, Ouni, Wahaishi, Desell, and
  Soui]{alshoaibi:22}
\BIBentryALTinterwordspacing
D.~Alshoaibi, M.~W. Mkaouer, A.~Ouni, A.~Wahaishi, T.~Desell, and M.~Soui,
  ``Search-based detection of code changes introducing performance
  regression,'' \emph{Swarm and Evolutionary Computation}, vol.~73, p. 101101,
  Aug. 2022. [Online]. Available:
  \url{https://doi.org/10.1016/j.swevo.2022.101101}
\BIBentrySTDinterwordspacing

\bibitem[Arcuri(2019)]{arcuri:19}
\BIBentryALTinterwordspacing
A.~Arcuri, ``{REST}ful {API} automated test case generation with
  {E}vo{M}aster,'' \emph{{ACM} Transactions on Software Engineering and
  Methodology}, vol.~28, no.~1, pp. 1--37, Feb. 2019. [Online]. Available:
  \url{https://doi.org/10.1145/3293455}
\BIBentrySTDinterwordspacing

\bibitem[Arcuri and Briand(2011)]{arcuri:11}
\BIBentryALTinterwordspacing
A.~Arcuri and L.~Briand, ``A practical guide for using statistical tests to
  assess randomized algorithms in software engineering,'' in \emph{Proceedings
  of the 33rd International Conference on Software Engineering}, ser. {ICSE}
  2011.\hskip 1em plus 0.5em minus 0.4em\relax {ACM}, 2011, pp. 1--10.
  [Online]. Available: \url{https://doi.org/10.1145/1985793.1985795}
\BIBentrySTDinterwordspacing

\bibitem[Auger et~al.(2009)Auger, Bader, Brockhoff, and Zitzler]{auger:09}
\BIBentryALTinterwordspacing
A.~Auger, J.~Bader, D.~Brockhoff, and E.~Zitzler, ``Theory of the hypervolume
  indicator: {O}ptimal {\textmu}-distributions and the choice of the reference
  point,'' in \emph{Proceedings of the 10th {ACM} {SIGEVO} Workshop on
  Foundations of Genetic Algorithms}, ser. {FOGA} 2009.\hskip 1em plus 0.5em
  minus 0.4em\relax {ACM}, 2009, pp. 87--102. [Online]. Available:
  \url{https://doi.org/10.1145/1527125.1527138}
\BIBentrySTDinterwordspacing

\bibitem[Benjamini and Yekutieli(2001)]{benjamini:01}
\BIBentryALTinterwordspacing
Y.~Benjamini and D.~Yekutieli, ``The control of the false discovery rate in
  multiple testing under dependency,'' \emph{The Annals of Statistics},
  vol.~29, no.~4, pp. 1165--1188, Aug. 2001. [Online]. Available:
  \url{https://doi.org/10.1214/aos/1013699998}
\BIBentrySTDinterwordspacing

\bibitem[Branke et~al.(2004)Branke, Deb, Dierolf, and Osswald]{branke:04}
\BIBentryALTinterwordspacing
J.~Branke, K.~Deb, H.~Dierolf, and M.~Osswald, ``Finding knees in
  multi-objective optimization,'' in \emph{Proceedings of the 8th International
  Conference on Parallel Problem Solving from Nature}, ser. {PPSN} 2004.\hskip
  1em plus 0.5em minus 0.4em\relax Springer, 2004, pp. 722--731. [Online].
  Available: \url{https://doi.org/10.1007/978-3-540-30217-9_73}
\BIBentrySTDinterwordspacing

\bibitem[Chen and Shang(2017)]{chen:17}
\BIBentryALTinterwordspacing
J.~Chen and W.~Shang, ``An exploratory study of performance regression
  introducing code changes,'' in \emph{Proceedings of the 33rd {IEEE}
  International Conference on Software Maintenance and Evolution}, ser. {ICSME}
  2017.\hskip 1em plus 0.5em minus 0.4em\relax New York, NY, USA: {IEEE}, Sep.
  2017, pp. 341--352. [Online]. Available:
  \url{https://doi.org/10.1109/icsme.2017.13}
\BIBentrySTDinterwordspacing

\bibitem[Chen et~al.(2022)Chen, Shang, and Shihab]{chen:22}
\BIBentryALTinterwordspacing
J.~Chen, W.~Shang, and E.~Shihab, ``{PerfJIT}: Test-level just-in-time
  prediction for performance regression introducing commits,'' \emph{{IEEE}
  Transactions on Software Engineering}, vol.~48, no.~5, pp. 1529--1544, May
  2022. [Online]. Available: \url{https://doi.org/10.1109%2Ftse.2020.3023955}
\BIBentrySTDinterwordspacing

\bibitem[Chen et~al.(2021)Chen, Li, Chen, Fan, Hu, and Yang]{chen:21}
\BIBentryALTinterwordspacing
K.~Chen, Y.~Li, Y.~Chen, C.~Fan, Z.~Hu, and W.~Yang, ``{GLIB}: Towards
  automated test oracle for graphically-rich applications,'' in
  \emph{Proceedings of the 29th {ACM} Joint European Software Engineering
  Conference and Symposium on the Foundations of Software Engineering}, ser.
  {ESEC}/{FSE} 2021.\hskip 1em plus 0.5em minus 0.4em\relax {ACM}, Aug. 2021,
  pp. 1093--1104. [Online]. Available:
  \url{https://doi.org/10.1145/3468264.3468586}
\BIBentrySTDinterwordspacing

\bibitem[Daly(2021)]{daly:21}
\BIBentryALTinterwordspacing
D.~Daly, ``Creating a virtuous cycle in performance testing at {MongoDB},'' in
  \emph{Proceedings of the 12th {ACM}/{SPEC} International Conference on
  Performance Engineering}, ser. {ICPE} 2021.\hskip 1em plus 0.5em minus
  0.4em\relax {ACM}, Apr. 2021. [Online]. Available:
  \url{https://doi.org/10.1145/3427921.3450234}
\BIBentrySTDinterwordspacing

\bibitem[Daly et~al.(2020)Daly, Brown, Ingo, O{\textquotesingle}Leary, and
  Bradford]{daly:20}
\BIBentryALTinterwordspacing
D.~Daly, W.~Brown, H.~Ingo, J.~O{\textquotesingle}Leary, and D.~Bradford, ``The
  use of change point detection to identify software performance regressions in
  a continuous integration system,'' in \emph{Proceedings of the 11th
  {ACM}/{SPEC} International Conference on Performance Engineering}, ser.
  {ICPE} 2020.\hskip 1em plus 0.5em minus 0.4em\relax {ACM}, Apr. 2020.
  [Online]. Available: \url{https://doi.org/10.1145/3358960.3375791}
\BIBentrySTDinterwordspacing

\bibitem[Damasceno~Costa et~al.(2021)Damasceno~Costa, Bezemer, Leitner, and
  Andrzejak]{costa:21}
\BIBentryALTinterwordspacing
D.~E. Damasceno~Costa, C.-P. Bezemer, P.~Leitner, and A.~Andrzejak, ``What's
  wrong with my benchmark results? {S}tudying bad practices in {JMH}
  benchmarks,'' \emph{{IEEE} Transactions on Software Engineering}, vol.~47,
  no.~7, pp. 1452--1467, Jul. 2021. [Online]. Available:
  \url{https://doi.org/10.1109/TSE.2019.2925345}
\BIBentrySTDinterwordspacing

\bibitem[Davison and Hinkley(1997)]{davison:97}
\BIBentryALTinterwordspacing
A.~C. Davison and D.~Hinkley, \emph{Bootstrap Methods and Their
  Application}.\hskip 1em plus 0.5em minus 0.4em\relax Cambridge University
  Press, Oct. 1997. [Online]. Available:
  \url{https://doi.org/10.1017/CBO9780511802843}
\BIBentrySTDinterwordspacing

\bibitem[de~Oliveira et~al.(2017)de~Oliveira, Fischmeister, Diwan, Hauswirth,
  and Sweeney]{oliveira:17}
\BIBentryALTinterwordspacing
A.~B. de~Oliveira, S.~Fischmeister, A.~Diwan, M.~Hauswirth, and P.~F. Sweeney,
  ``{Perphecy}: Performance regression test selection made simple but
  effective,'' in \emph{Proceedings of the 10th {IEEE} International Conference
  on Software Testing, Verification and Validation}, ser. {ICST} 2017, Mar.
  2017, pp. 103--113. [Online]. Available:
  \url{https://doi.org/10.1109/ICST.2017.17}
\BIBentrySTDinterwordspacing

\bibitem[Deb and Jain(2014)]{deb:14}
\BIBentryALTinterwordspacing
K.~Deb and H.~Jain, ``An evolutionary many-objective optimization algorithm
  using reference-point-based nondominated sorting approach, part {I}: Solving
  problems with box constraints,'' \emph{{IEEE} Transactions on Evolutionary
  Computation}, vol.~18, no.~4, pp. 577--601, Aug. 2014. [Online]. Available:
  \url{https://doi.org/10.1109%2Ftevc.2013.2281535}
\BIBentrySTDinterwordspacing

\bibitem[Deb et~al.(2002)Deb, Pratap, Agarwal, and Meyarivan]{deb:02}
\BIBentryALTinterwordspacing
K.~Deb, A.~Pratap, S.~Agarwal, and T.~Meyarivan, ``A fast and elitist
  multiobjective genetic algorithm: {NSGA}-{II},'' \emph{{IEEE} Transactions on
  Evolutionary Computation}, vol.~6, no.~2, pp. 182--197, Apr. 2002. [Online].
  Available: \url{https://doi.org/10.1109%2F4235.996017}
\BIBentrySTDinterwordspacing

\bibitem[Delgado-P{\'e}rez et~al.(2020)Delgado-P{\'e}rez, S{\'a}nchez, Segura,
  and Medina-Bulo]{delgado-perez:20}
\BIBentryALTinterwordspacing
P.~Delgado-P{\'e}rez, A.~B. S{\'a}nchez, S.~Segura, and I.~Medina-Bulo,
  ``Performance mutation testing,'' \emph{Software Testing, Verification and
  Reliability}, vol.~31, no.~5, p. e1728, Jan. 2020. [Online]. Available:
  \url{https://doi.org/10.1002/stvr.1728}
\BIBentrySTDinterwordspacing

\bibitem[Devroey et~al.(2023)Devroey, Gambi, Galeotti, Just, Kifetew,
  Panichella, and Panichella]{devroey:23}
\BIBentryALTinterwordspacing
X.~Devroey, A.~Gambi, J.~P. Galeotti, R.~Just, F.~M. Kifetew, A.~Panichella,
  and S.~Panichella, ``{JUGE}: An infrastructure for benchmarking {J}ava unit
  test generators,'' \emph{Software Testing, Verification and Reliability},
  vol.~33, no.~3, p. e1838, May 2023. [Online]. Available:
  \url{https://doi.org/10.1002/stvr.1838}
\BIBentrySTDinterwordspacing

\bibitem[{Di Nucci} et~al.(2020){Di Nucci}, Panichella, Zaidman, and {De
  Lucia}]{dinucci:20}
\BIBentryALTinterwordspacing
D.~{Di Nucci}, A.~Panichella, A.~Zaidman, and A.~{De Lucia}, ``A test case
  prioritization genetic algorithm guided by the hypervolume indicator,''
  \emph{{IEEE} Transactions on Software Engineering}, vol.~46, no.~6, pp.
  674--696, Jun. 2020. [Online]. Available:
  \url{https://doi.org/10.1109/tse.2018.2868082}
\BIBentrySTDinterwordspacing

\bibitem[Dunn(1964)]{dunn:64}
\BIBentryALTinterwordspacing
O.~J. Dunn, ``Multiple comparisons using rank sums,'' \emph{Technometrics},
  vol.~6, no.~3, pp. 241--252, Aug. 1964. [Online]. Available:
  \url{https://doi.org/10.1080/00401706.1964.10490181}
\BIBentrySTDinterwordspacing

\bibitem[Elbaum et~al.(2001)Elbaum, Malishevsky, and Rothermel]{elbaum:01}
\BIBentryALTinterwordspacing
S.~Elbaum, A.~Malishevsky, and G.~Rothermel, ``Incorporating varying test costs
  and fault severities into test case prioritization,'' in \emph{Proceedings of
  the 23rd International Conference on Software Engineering}, ser. {ICSE}
  2001.\hskip 1em plus 0.5em minus 0.4em\relax Washington, DC, USA: {IEEE},
  2001, pp. 329--338. [Online]. Available:
  \url{https://doi.org/10.1109/icse.2001.919106}
\BIBentrySTDinterwordspacing

\bibitem[Elbaum et~al.(2014)Elbaum, Rothermel, and Penix]{elbaum:14}
\BIBentryALTinterwordspacing
S.~Elbaum, G.~Rothermel, and J.~Penix, ``Techniques for improving regression
  testing in continuous integration development environments,'' in
  \emph{Proceedings of the 22nd {ACM} {SIGSOFT} International Symposium on
  Foundations of Software Engineering}, ser. {FSE} 2014.\hskip 1em plus 0.5em
  minus 0.4em\relax New York, NY, USA: {ACM}, 2014, pp. 235--245. [Online].
  Available: \url{http://doi.acm.org/10.1145/2635868.2635910}
\BIBentrySTDinterwordspacing

\bibitem[Elsner et~al.(2021)Elsner, Hauer, Pretschner, and Reimer]{elsner:21}
\BIBentryALTinterwordspacing
D.~Elsner, F.~Hauer, A.~Pretschner, and S.~Reimer, ``Empirically evaluating
  readily available information for regression test optimization in continuous
  integration,'' in \emph{Proceedings of the 30th {ACM} {SIGSOFT} International
  Symposium on Software Testing and Analysis}, ser. {ISSTA} 2021.\hskip 1em
  plus 0.5em minus 0.4em\relax {ACM}, Jul. 2021, pp. 491--504. [Online].
  Available: \url{https://doi.org/10.1145/3460319.3464834}
\BIBentrySTDinterwordspacing

\bibitem[Epitropakis et~al.(2015)Epitropakis, Yoo, Harman, and
  Burke]{epitropakis:15}
\BIBentryALTinterwordspacing
M.~G. Epitropakis, S.~Yoo, M.~Harman, and E.~K. Burke, ``Empirical evaluation
  of pareto efficient multi-objective regression test case prioritisation,'' in
  \emph{Proceedings of the 2015 International Symposium on Software Testing and
  Analysis}, ser. {ISSTA} 2015.\hskip 1em plus 0.5em minus 0.4em\relax {ACM},
  Jul. 2015, pp. 234--245. [Online]. Available:
  \url{https://doi.org/10.1145/2771783.2771788}
\BIBentrySTDinterwordspacing

\bibitem[Fonseca and Fleming(1995)]{fonseca:95}
\BIBentryALTinterwordspacing
C.~M. Fonseca and P.~J. Fleming, ``An overview of evolutionary algorithms in
  multiobjective optimization,'' \emph{Evolutionary Computation}, vol.~3,
  no.~1, pp. 1--16, Mar. 1995. [Online]. Available:
  \url{https://doi.org/10.1162/evco.1995.3.1.1}
\BIBentrySTDinterwordspacing

\bibitem[Fraser and Zeller(2010)]{fraser:10}
\BIBentryALTinterwordspacing
G.~Fraser and A.~Zeller, ``Mutation-driven generation of unit tests and
  oracles,'' in \emph{Proceedings of the 19th International Symposium on
  Software Testing and Analysis}, ser. {ISSTA} 2010.\hskip 1em plus 0.5em minus
  0.4em\relax {ACM}, 2010, pp. 147--158. [Online]. Available:
  \url{https://doi.org/10.1145/1831708.1831728}
\BIBentrySTDinterwordspacing

\bibitem[Georges et~al.(2007)Georges, Buytaert, and Eeckhout]{georges:07}
\BIBentryALTinterwordspacing
A.~Georges, D.~Buytaert, and L.~Eeckhout, ``Statistically rigorous {J}ava
  performance evaluation,'' in \emph{Proceedings of the 22nd {ACM} {SIGPLAN}
  Conference on Object-Oriented Programming, Systems, and Applications}, ser.
  {OOPSLA} 2007.\hskip 1em plus 0.5em minus 0.4em\relax New York, NY, USA:
  {ACM}, Oct. 2007, pp. 57--76. [Online]. Available:
  \url{http://doi.acm.org/10.1145/1297027.1297033}
\BIBentrySTDinterwordspacing

\bibitem[Grambow et~al.(2021)Grambow, Laaber, Leitner, and
  Bermbach]{grambow:21}
\BIBentryALTinterwordspacing
M.~Grambow, C.~Laaber, P.~Leitner, and D.~Bermbach, ``Using application
  benchmark call graphs to quantify and improve the practical relevance of
  microbenchmark suites,'' \emph{{PeerJ} Computer Science}, vol.~7, no. e548,
  pp. 1--32, May 2021. [Online]. Available:
  \url{https://doi.org/10.7717/peerj-cs.548}
\BIBentrySTDinterwordspacing

\bibitem[Grambow et~al.(2023)Grambow, Kovalev, Laaber, Leitner, and
  Bermbach]{grambow:23}
M.~Grambow, D.~Kovalev, C.~Laaber, P.~Leitner, and D.~Bermbach, ``Using
  microbenchmark suites to detect application performance changes,''
  \emph{{IEEE} Transactions on Cloud Computing}, vol.~11, no.~3, pp.
  2575--2590, Jul. 2023.

\bibitem[Haghighatkhah et~al.(2018)Haghighatkhah, M{\"a}ntyl{\"a}, Oivo, and
  Kuvaja]{haghighatkhah:18}
\BIBentryALTinterwordspacing
A.~Haghighatkhah, M.~M{\"a}ntyl{\"a}, M.~Oivo, and P.~Kuvaja, ``Test
  prioritization in continuous integration environments,'' \emph{The Journal of
  Systems and Software}, vol. 146, pp. 80--98, Dec. 2018. [Online]. Available:
  \url{https://doi.org/10.1016/j.jss.2018.08.061}
\BIBentrySTDinterwordspacing

\bibitem[Hao et~al.(2014)Hao, Zhang, Zhang, Rothermel, and Mei]{hao:14}
\BIBentryALTinterwordspacing
D.~Hao, L.~Zhang, L.~Zhang, G.~Rothermel, and H.~Mei, ``A unified test case
  prioritization approach,'' \emph{{ACM} Transactions on Software Engineering
  and Methodology}, vol.~24, no.~2, pp. 1--31, Dec. 2014. [Online]. Available:
  \url{http://doi.acm.org/10.1145/2685614}
\BIBentrySTDinterwordspacing

\bibitem[Harman(2007)]{harman:07}
\BIBentryALTinterwordspacing
M.~Harman, ``The current state and future of search based software
  engineering,'' in \emph{Future of Software Engineering}, ser. {FOSE}
  2007.\hskip 1em plus 0.5em minus 0.4em\relax {IEEE}, May 2007, pp. 342--357.
  [Online]. Available: \url{https://doi.org/10.1109/fose.2007.29}
\BIBentrySTDinterwordspacing

\bibitem[Harman and Jones(2001)]{harman:01}
\BIBentryALTinterwordspacing
M.~Harman and B.~F. Jones, ``Search-based software engineering,''
  \emph{Information and Software Technology}, vol.~43, no.~14, pp. 833--839,
  Dec. 2001. [Online]. Available:
  \url{https://doi.org/10.1016/S0950-5849(01)00189-6}
\BIBentrySTDinterwordspacing

\bibitem[He et~al.(2019)He, Manns, Saunders, Wang, Pollock, and Soffa]{he:19}
\BIBentryALTinterwordspacing
S.~He, G.~Manns, J.~Saunders, W.~Wang, L.~Pollock, and M.~L. Soffa, ``A
  statistics-based performance testing methodology for cloud applications,'' in
  \emph{Proceedings of the 27th {ACM} Joint European Software Engineering
  Conference and Symposium on the Foundations of Software Engineering}, ser.
  {ESEC}/{FSE} 2019.\hskip 1em plus 0.5em minus 0.4em\relax New York, NY, USA:
  {ACM}, Aug. 2019, pp. 188--199. [Online]. Available:
  \url{http://doi.acm.org/10.1145/3338906.3338912}
\BIBentrySTDinterwordspacing

\bibitem[Hess and Kromrey(2004)]{hess:04}
M.~R. Hess and J.~D. Kromrey, ``Robust confidence intervals for effect sizes: A
  comparative study of cohen's d and cliff's delta under non-normality and
  heterogeneous variances,'' \emph{Annual Meeting of the American Educational
  Research Association}, Apr. 2004.

\bibitem[Hesterberg(2015)]{hesterberg:15}
\BIBentryALTinterwordspacing
T.~C. Hesterberg, ``What teachers should know about the bootstrap: Resampling
  in the undergraduate statistics curriculum,'' \emph{The American
  Statistician}, vol.~69, no.~4, pp. 371--386, 2015. [Online]. Available:
  \url{https://doi.org/10.1080/00031305.2015.1089789}
\BIBentrySTDinterwordspacing

\bibitem[Huang et~al.(2014)Huang, Ma, Shen, and Zhou]{huang:14}
\BIBentryALTinterwordspacing
P.~Huang, X.~Ma, D.~Shen, and Y.~Zhou, ``Performance regression testing target
  prioritization via performance risk analysis,'' in \emph{Proceedings of the
  36th {IEEE}/{ACM} International Conference on Software Engineering}, ser.
  {ICSE} 2014.\hskip 1em plus 0.5em minus 0.4em\relax New York, NY, USA: {ACM},
  May 2014, pp. 60--71. [Online]. Available:
  \url{http://doi.acm.org/10.1145/2568225.2568232}
\BIBentrySTDinterwordspacing

\bibitem[Islam et~al.(2012)Islam, Marchetto, Susi, and Scanniello]{islam:12}
\BIBentryALTinterwordspacing
M.~M. Islam, A.~Marchetto, A.~Susi, and G.~Scanniello, ``A multi-objective
  technique to prioritize test cases based on latent semantic indexing,'' in
  \emph{Proceedings of the 16th European Conference on Software Maintenance and
  Reengineering}, ser. {CSMR} 2012.\hskip 1em plus 0.5em minus 0.4em\relax
  {IEEE}, Mar. 2012, pp. 21--30. [Online]. Available:
  \url{https://doi.org/10.1109/csmr.2012.13}
\BIBentrySTDinterwordspacing

\bibitem[Jangali et~al.(2022)Jangali, Tang, Alexandersson, Leitner, Yang, and
  Shang]{jangali:22}
\BIBentryALTinterwordspacing
M.~Jangali, Y.~Tang, N.~Alexandersson, P.~Leitner, J.~Yang, and W.~Shang,
  ``Automated generation and evaluation of {JMH} microbenchmark suites from
  unit tests,'' \emph{{IEEE} Transactions on Software Engineering}, pp. 1--23,
  2022. [Online]. Available: \url{https://doi.org/10.1109/TSE.2022.3188005}
\BIBentrySTDinterwordspacing

\bibitem[Jiang and Hassan(2015)]{jiang:15}
\BIBentryALTinterwordspacing
Z.~M. Jiang and A.~E. Hassan, ``A survey on load testing of large-scale
  software systems,'' \emph{{IEEE} Transactions on Software Engineering},
  vol.~41, no.~11, pp. 1091--1118, Nov. 2015. [Online]. Available:
  \url{https://doi.org/10.1109/tse.2015.2445340}
\BIBentrySTDinterwordspacing

\bibitem[Just et~al.(2014)Just, Jalali, and Ernst]{just:14}
\BIBentryALTinterwordspacing
R.~Just, D.~Jalali, and M.~D. Ernst, ``{Defects4J}: {A} database of existing
  faults to enable controlled testing studies for {J}ava programs,'' in
  \emph{Proceedings of the 2014 International Symposium on Software Testing and
  Analysis}, ser. {ISSTA} 2014.\hskip 1em plus 0.5em minus 0.4em\relax {ACM},
  2014, pp. 437--440. [Online]. Available:
  \url{https://doi.org/10.1145/2610384.2628055}
\BIBentrySTDinterwordspacing

\bibitem[Kalibera and Jones(2012)]{kalibera:12}
\BIBentryALTinterwordspacing
T.~Kalibera and R.~Jones, ``Quantifying performance changes with effect size
  confidence intervals,'' University of Kent, Technical Report 4--12, Jun.
  2012, accessed: 10.11.2022. [Online]. Available:
  \url{http://www.cs.kent.ac.uk/pubs/2012/3233}
\BIBentrySTDinterwordspacing

\bibitem[Kalibera and Jones(2013)]{kalibera:13}
\BIBentryALTinterwordspacing
------, ``Rigorous benchmarking in reasonable time,'' in \emph{Proceedings of
  the 2013 {ACM} {SIGPLAN} International Symposium on Memory Management}, ser.
  {ISMM} 2013.\hskip 1em plus 0.5em minus 0.4em\relax New York, NY, USA: {ACM},
  2013, pp. 63--74. [Online]. Available:
  \url{http://doi.acm.org/10.1145/2464157.2464160}
\BIBentrySTDinterwordspacing

\bibitem[Kim et~al.(2004)Kim, Hiroyasu, Miki, and Watanabe]{kim:04}
\BIBentryALTinterwordspacing
M.~Kim, T.~Hiroyasu, M.~Miki, and S.~Watanabe, ``{SPEA}2+: {I}mproving the
  performance of the strength pareto evolutionary algorithm 2,'' in
  \emph{Proceedings of the 8th International Conference on Parallel Problem
  Solving from Nature}, ser. {PPSN} 2004, vol. 3242.\hskip 1em plus 0.5em minus
  0.4em\relax Springer, 2004, pp. 742--751. [Online]. Available:
  \url{https://doi.org/10.1007%2F978-3-540-30217-9_75}
\BIBentrySTDinterwordspacing

\bibitem[Knowles and Corne(1999)]{knowles:99}
\BIBentryALTinterwordspacing
J.~D. Knowles and D.~Corne, ``The {P}areto aarchived evolution strategy: {A}
  new baseline algorithm for {P}areto {M}ultiobjective optimisation,'' in
  \emph{Proceedings of the Congress on Evolutionary Computation}, ser. {CEC}
  1999.\hskip 1em plus 0.5em minus 0.4em\relax {IEEE}, Jul. 1999, pp. 98--105.
  [Online]. Available: \url{https://doi.org/10.1109/cec.1999.781913}
\BIBentrySTDinterwordspacing

\bibitem[Kruskal and Wallis(1952)]{kruskal:52}
\BIBentryALTinterwordspacing
W.~H. Kruskal and W.~A. Wallis, ``Use of ranks in one-criterion variance
  analysis,'' \emph{Journal of the American Statistical Association}, vol.~47,
  no. 260, pp. 583--621, Dec. 1952. [Online]. Available:
  \url{https://doi.org/10.1080/01621459.1952.10483441}
\BIBentrySTDinterwordspacing

\bibitem[Laaber(2022)]{laaber:pa-v0.1.0-zenodo}
\BIBentryALTinterwordspacing
C.~Laaber, ``chrstphlbr/pa: v0.1.0,'' Nov. 2022. [Online]. Available:
  \url{https://doi.org/10.5281/zenodo.7308066}
\BIBentrySTDinterwordspacing

\bibitem[Laaber and Leitner(2018)]{laaber:18}
\BIBentryALTinterwordspacing
C.~Laaber and P.~Leitner, ``An evaluation of open-source software
  microbenchmark suites for continuous performance assessment,'' in
  \emph{Proceedings of the 15th International Conference on Mining Software
  Repositories}, ser. {MSR} 2018.\hskip 1em plus 0.5em minus 0.4em\relax New
  York, NY, USA: {ACM}, 2018, pp. 119--130. [Online]. Available:
  \url{http://doi.acm.org/10.1145/3196398.3196407}
\BIBentrySTDinterwordspacing

\bibitem[Laaber and W{\"u}rsten(2024)]{laaber:bencher-v0.4.0-zenodo}
\BIBentryALTinterwordspacing
C.~Laaber and S.~W{\"u}rsten, ``chrstphlbr/bencher: Release v0.4.0,'' Jan.
  2024. [Online]. Available: \url{https://doi.org/10.5281/zenodo.10527360}
\BIBentrySTDinterwordspacing

\bibitem[Laaber et~al.(2019)Laaber, Scheuner, and Leitner]{laaber:19}
\BIBentryALTinterwordspacing
C.~Laaber, J.~Scheuner, and P.~Leitner, ``Software microbenchmarking in the
  cloud. {H}ow bad is it really?'' \emph{Empirical Software Engineering},
  vol.~24, pp. 2469--2508, Aug. 2019. [Online]. Available:
  \url{https://doi.org/10.1007/s10664-019-09681-1}
\BIBentrySTDinterwordspacing

\bibitem[Laaber et~al.(2020)Laaber, W{\"u}rsten, Gall, and Leitner]{laaber:20}
\BIBentryALTinterwordspacing
C.~Laaber, S.~W{\"u}rsten, H.~C. Gall, and P.~Leitner, ``Dynamically
  reconfiguring software microbenchmarks: Reducing execution time without
  sacrificing result quality,'' in \emph{Proceedings of the 28th {ACM} Joint
  European Software Engineering Conference and Symposium on the Foundations of
  Software Engineering}, ser. {ESEC}/{FSE} 2020.\hskip 1em plus 0.5em minus
  0.4em\relax {ACM}, Nov. 2020, pp. 989--1001. [Online]. Available:
  \url{https://doi.org/10.1145/3368089.3409683}
\BIBentrySTDinterwordspacing

\bibitem[Laaber et~al.(2021{\natexlab{c}})Laaber, Basmaci, and
  Salza]{laaber:21b}
\BIBentryALTinterwordspacing
C.~Laaber, M.~Basmaci, and P.~Salza, ``Predicting unstable software benchmarks
  using static source code features,'' \emph{Empirical Software Engineering},
  vol.~26, no.~6, pp. 1--53, Aug. 2021. [Online]. Available:
  \url{https://doi.org/10.1007/s10664-021-09996-y}
\BIBentrySTDinterwordspacing

\bibitem[Laaber et~al.(2021{\natexlab{a}})Laaber, Gall, and
  Leitner]{laaber:21c}
\BIBentryALTinterwordspacing
C.~Laaber, H.~C. Gall, and P.~Leitner, ``Applying test case prioritization to
  software microbenchmarks,'' \emph{Empirical Software Engineering}, vol.~26,
  no.~6, pp. 1--48, Sep. 2021. [Online]. Available:
  \url{https://doi.org/10.1007/s10664-021-10037-x}
\BIBentrySTDinterwordspacing

\bibitem[Laaber et~al.(2021{\natexlab{b}})Laaber, Gall, and
  Leitner]{laaber:21c:replication_package}
\BIBentryALTinterwordspacing
------, ``Replication package "{A}pplying test case prioritization to software
  microbenchmarks",'' 2021. [Online]. Available:
  \url{https://doi.org/10.5281/zenodo.5206117}
\BIBentrySTDinterwordspacing

\bibitem[Laaber et~al.(2024)Laaber, Yue, and
  Ali]{laaber:24:replication_package_v0_2_0}
\BIBentryALTinterwordspacing
C.~Laaber, T.~Yue, and S.~Ali, ``Replication package "{E}valuating search-based
  software microbenchmark prioritization",'' 2024. [Online]. Available:
  \url{https://doi.org/10.5281/zenodo.10527125}
\BIBentrySTDinterwordspacing

\bibitem[Leitner and Bezemer(2017)]{leitner:17}
\BIBentryALTinterwordspacing
P.~Leitner and C.-P. Bezemer, ``An exploratory study of the state of practice
  of performance testing in {J}ava-based open source projects,'' in
  \emph{Proceedings of the 8th {ACM}/{SPEC} on International Conference on
  Performance Engineering}, ser. {ICPE} 2017.\hskip 1em plus 0.5em minus
  0.4em\relax New York, NY, USA: {ACM}, 2017, pp. 373--384. [Online].
  Available: \url{http://doi.acm.org/10.1145/3030207.3030213}
\BIBentrySTDinterwordspacing

\bibitem[Li et~al.(2007)Li, Harman, and Hierons]{li:07}
\BIBentryALTinterwordspacing
Z.~Li, M.~Harman, and R.~M. Hierons, ``Search algorithms for regression test
  case prioritization,'' \emph{{IEEE} Transactions on Software Engineering},
  vol.~33, no.~4, pp. 225--237, Apr. 2007. [Online]. Available:
  \url{https://doi.org/10.1109/TSE.2007.38}
\BIBentrySTDinterwordspacing

\bibitem[Li et~al.(2013)Li, Bian, Zhao, and Cheng]{li:13}
\BIBentryALTinterwordspacing
Z.~Li, Y.~Bian, R.~Zhao, and J.~Cheng, ``A fine-grained parallel
  multi-objective test case prioritization on {GPU},'' in \emph{Proceedings of
  the 5th Symposium on Search Based Software Engineering}, ser. {SSBSE}
  2013.\hskip 1em plus 0.5em minus 0.4em\relax Springer, 2013, pp. 111--125.
  [Online]. Available: \url{https://doi.org/10.1007/978-3-642-39742-4_10}
\BIBentrySTDinterwordspacing

\bibitem[Liang et~al.(2018)Liang, Elbaum, and Rothermel]{liang:18}
\BIBentryALTinterwordspacing
J.~Liang, S.~Elbaum, and G.~Rothermel, ``Redefining prioritization: Continuous
  prioritization for continuous integration,'' in \emph{Proceedings of the 40th
  {IEEE}/{ACM} International Conference on Software Engineering}, ser. {ICSE}
  2018.\hskip 1em plus 0.5em minus 0.4em\relax New York, NY, USA: {ACM}, May
  2018, pp. 688--698. [Online]. Available:
  \url{http://doi.acm.org/10.1145/3180155.3180213}
\BIBentrySTDinterwordspacing

\bibitem[Luo et~al.(2016)Luo, Moran, and Poshyvanyk]{luo:16a}
\BIBentryALTinterwordspacing
Q.~Luo, K.~Moran, and D.~Poshyvanyk, ``A large-scale empirical comparison of
  static and dynamic test case prioritization techniques,'' in
  \emph{Proceedings of the 24th {ACM} {SIGSOFT} International Symposium on
  Foundations of Software Engineering}, ser. {FSE} 2016.\hskip 1em plus 0.5em
  minus 0.4em\relax New York, NY, USA: {ACM}, Nov. 2016, pp. 559--570.
  [Online]. Available: \url{http://doi.acm.org/10.1145/2950290.2950344}
\BIBentrySTDinterwordspacing

\bibitem[Luo et~al.(2019)Luo, Moran, Zhang, and Poshyvanyk]{luo:19}
\BIBentryALTinterwordspacing
Q.~Luo, K.~Moran, L.~Zhang, and D.~Poshyvanyk, ``How do static and dynamic test
  case prioritization techniques perform on modern software systems? {A}n
  extensive study on {GitHub} projects,'' \emph{{IEEE} Transactions on Software
  Engineering}, vol.~45, no.~11, pp. 1054--1080, Nov. 2019. [Online].
  Available: \url{https://doi.org/10.1109/tse.2018.2822270}
\BIBentrySTDinterwordspacing

\bibitem[Marchetto et~al.(2016)Marchetto, Islam, Asghar, Susi, and
  Scanniello]{marchetto:16}
\BIBentryALTinterwordspacing
A.~Marchetto, M.~M. Islam, W.~Asghar, A.~Susi, and G.~Scanniello, ``A
  multi-objective technique to prioritize test cases,'' \emph{{IEEE}
  Transactions on Software Engineering}, vol.~42, no.~10, pp. 918--940, Oct.
  2016. [Online]. Available: \url{https://doi.org/10.1109/tse.2015.2510633}
\BIBentrySTDinterwordspacing

\bibitem[Maricq et~al.(2018)Maricq, Duplyakin, Jimenez, Maltzahn, Stutsman, and
  Ricci]{maricq:18}
\BIBentryALTinterwordspacing
A.~Maricq, D.~Duplyakin, I.~Jimenez, C.~Maltzahn, R.~Stutsman, and R.~Ricci,
  ``Taming performance variability,'' in \emph{Proceedings of the 13th {USENIX}
  Conference on Operating Systems Design and Implementation}, ser. {OSDI}
  2018.\hskip 1em plus 0.5em minus 0.4em\relax USA: {USENIX} Association, Oct.
  2018, pp. 409--425. [Online]. Available:
  \url{https://www.usenix.org/conference/osdi18/presentation/maricq}
\BIBentrySTDinterwordspacing

\bibitem[Mostafa et~al.(2017)Mostafa, Wang, and Xie]{mostafa:17}
\BIBentryALTinterwordspacing
S.~Mostafa, X.~Wang, and T.~Xie, ``{PerfRanker}: Prioritization of performance
  regression tests for collection-intensive software,'' in \emph{Proceedings of
  the 26th {ACM} {SIGSOFT} International Symposium on Software Testing and
  Analysis}, ser. {ISSTA} 2017.\hskip 1em plus 0.5em minus 0.4em\relax New
  York, NY, USA: {ACM}, 2017, pp. 23--34. [Online]. Available:
  \url{http://doi.acm.org/10.1145/3092703.3092725}
\BIBentrySTDinterwordspacing

\bibitem[Nebro et~al.(2009)Nebro, Durillo, Luna, Dorronsoro, and
  Alba]{nebro:09}
\BIBentryALTinterwordspacing
A.~J. Nebro, J.~J. Durillo, F.~Luna, B.~Dorronsoro, and E.~Alba, ``{MOCell}:
  {A} cellular genetic algorithm for multiobjective optimization,''
  \emph{International Journal of Intelligent Systems}, vol.~24, no.~7, pp.
  726--746, Jul. 2009. [Online]. Available:
  \url{https://doi.org/10.1002/int.20358}
\BIBentrySTDinterwordspacing

\bibitem[Nebro et~al.(2021)Nebro, P{\'e}rez-Abad, Aldana-Martin, and
  Garc{\'\i}a-Nieto]{nebro:21}
\BIBentryALTinterwordspacing
A.~J. Nebro, J.~P{\'e}rez-Abad, J.~F. Aldana-Martin, and J.~Garc{\'\i}a-Nieto,
  \emph{Evolving a Multi-Objective Optimization Framework}.\hskip 1em plus
  0.5em minus 0.4em\relax Springer, May 2021, pp. 175--198. [Online].
  Available: \url{https://doi.org/10.1007/978-981-16-0662-5_9}
\BIBentrySTDinterwordspacing

\bibitem[Pradel et~al.(2014)Pradel, Huggler, and Gross]{pradel:14}
\BIBentryALTinterwordspacing
M.~Pradel, M.~Huggler, and T.~R. Gross, ``Performance regression testing of
  concurrent classes,'' in \emph{Proceedings of the 2014 International
  Symposium on Software Testing and Analysis}, ser. {ISSTA} 2014.\hskip 1em
  plus 0.5em minus 0.4em\relax {ACM}, Jul. 2014, pp. 13--25. [Online].
  Available: \url{http://doi.acm.org/10.1145/2610384.2610393}
\BIBentrySTDinterwordspacing

\bibitem[Ren et~al.(2010)Ren, Lai, Tong, Aminzadeh, Hou, and Lai]{ren:10}
\BIBentryALTinterwordspacing
S.~Ren, H.~Lai, W.~Tong, M.~Aminzadeh, X.~Hou, and S.~Lai, ``Nonparametric
  bootstrapping for hierarchical data,'' \emph{Journal of Applied Statistics},
  vol.~37, no.~9, pp. 1487--1498, 2010. [Online]. Available:
  \url{https://doi.org/10.1080/02664760903046102}
\BIBentrySTDinterwordspacing

\bibitem[Rothermel and Harrold(1998)]{rothermel:98}
\BIBentryALTinterwordspacing
G.~Rothermel and M.~J. Harrold, ``Empirical studies of a safe regression test
  selection technique,'' \emph{{IEEE} Transactions on Software Engineering},
  vol.~24, no.~6, pp. 401--419, Jun. 1998. [Online]. Available:
  \url{http://doi.org/10.1109/32.689399}
\BIBentrySTDinterwordspacing

\bibitem[Rothermel et~al.(1999)Rothermel, Untch, Chu, and
  Harrold]{rothermel:99}
\BIBentryALTinterwordspacing
G.~Rothermel, R.~H. Untch, C.~Chu, and M.~J. Harrold, ``Test case
  prioritization: An empirical study,'' in \emph{Proceedings of the {IEEE}
  International Conference on Software Maintenance}, ser. {ICSM} 1999.\hskip
  1em plus 0.5em minus 0.4em\relax Washington, DC, USA: {IEEE}, 1999, pp.
  179--. [Online]. Available: \url{https://doi.org/10.1109/icsm.1999.792604}
\BIBentrySTDinterwordspacing

\bibitem[Rothermel et~al.(2001)Rothermel, Untch, and Chu]{rothermel:01}
\BIBentryALTinterwordspacing
G.~Rothermel, R.~J. Untch, and C.~Chu, ``Prioritizing test cases for regression
  testing,'' \emph{{IEEE} Transactions on Software Engineering}, vol.~27,
  no.~10, pp. 929--948, Oct. 2001. [Online]. Available:
  \url{https://doi.org/10.1109/32.962562}
\BIBentrySTDinterwordspacing

\bibitem[Samoaa and Leitner(2021)]{samoaa:21}
\BIBentryALTinterwordspacing
H.~Samoaa and P.~Leitner, ``An exploratory study of the impact of
  parameterization on {JMH} measurement results in open-source projects,'' in
  \emph{Proceedings of the 12th {ACM}/{SPEC} International Conference on
  Performance Engineering}, ser. {ICPE} 2021.\hskip 1em plus 0.5em minus
  0.4em\relax {ACM}, Apr. 2021, pp. 213--224. [Online]. Available:
  \url{https://doi.org/10.1145/3427921.3450243}
\BIBentrySTDinterwordspacing

\bibitem[Stefan et~al.(2017)Stefan, Hork{\'y}, Bulej, and T{\r u}ma]{stefan:17}
\BIBentryALTinterwordspacing
P.~Stefan, V.~Hork{\'y}, L.~Bulej, and P.~T{\r u}ma, ``Unit testing performance
  in {J}ava projects: Are we there yet?'' in \emph{Proceedings of the 8th
  {ACM}/{SPEC} on International Conference on Performance Engineering}, ser.
  {ICPE} 2017.\hskip 1em plus 0.5em minus 0.4em\relax New York, NY, USA: {ACM},
  Apr. 2017, pp. 401--412. [Online]. Available:
  \url{http://doi.acm.org/10.1145/3030207.3030226}
\BIBentrySTDinterwordspacing

\bibitem[Stol and Fitzgerald(2018)]{stol:18}
\BIBentryALTinterwordspacing
K.-J. Stol and B.~Fitzgerald, ``The {ABC} of software engineering research,''
  \emph{{ACM} Transactions on Software Engineering and Methodology}, vol.~27,
  no.~3, pp. 1--51, Oct. 2018. [Online]. Available:
  \url{https://doi.org/10.1145/3241743}
\BIBentrySTDinterwordspacing

\bibitem[Traini et~al.(2022)Traini, {Di Pompeo}, Tucci, Lin, Scalabrino,
  Bavota, Lanza, Oliveto, and Cortellessa]{traini:22b}
\BIBentryALTinterwordspacing
L.~Traini, D.~{Di Pompeo}, M.~Tucci, B.~Lin, S.~Scalabrino, G.~Bavota,
  M.~Lanza, R.~Oliveto, and V.~Cortellessa, ``How software refactoring impacts
  execution time,'' \emph{{ACM} Transactions on Software Engineering and
  Methodology}, vol.~31, no.~2, pp. 1--23, Apr. 2022. [Online]. Available:
  \url{https://doi.org/10.1145/3485136}
\BIBentrySTDinterwordspacing

\bibitem[Traini et~al.(2023)Traini, Cortellessa, {Di Pompeo}, and
  Tucci]{traini:23}
\BIBentryALTinterwordspacing
L.~Traini, V.~Cortellessa, D.~{Di Pompeo}, and M.~Tucci, ``Towards effective
  assessment of steady state performance in {J}ava software: Are we there
  yet?'' \emph{Empirical Software Engineering}, vol.~28, no.~13, pp. 1--57,
  2023. [Online]. Available: \url{https://doi.org/10.1007/s10664-022-10247-x}
\BIBentrySTDinterwordspacing

\bibitem[Vargha and Delaney(2000)]{vargha:00}
\BIBentryALTinterwordspacing
A.~Vargha and H.~D. Delaney, ``A critique and improvement of the "{CL}" common
  language effect size statistics of {McGraw} and {Wong},'' \emph{Journal of
  Educational and Behavioral Statistics}, vol.~25, no.~2, pp. 101--132, 2000.
  [Online]. Available: \url{https://doi.org/10.2307/1165329}
\BIBentrySTDinterwordspacing

\bibitem[Yoo and Harman(2012)]{yoo:12}
\BIBentryALTinterwordspacing
S.~Yoo and M.~Harman, ``Regression testing minimization, selection and
  prioritization: A survey,'' \emph{Software: Testing, Verification and
  Reliability}, vol.~22, no.~2, pp. 67--120, Mar. 2012. [Online]. Available:
  \url{https://doi.org/10.1002/stvr.430}
\BIBentrySTDinterwordspacing

\bibitem[Yoo and Harman(2007)]{yoo:07}
\BIBentryALTinterwordspacing
------, ``Pareto efficient multi-objective test case selection,'' in
  \emph{Proceedings of the 2007 International Symposium on Software Testing and
  Analysis}, ser. {ISSTA} 2007.\hskip 1em plus 0.5em minus 0.4em\relax {ACM},
  Jul. 2007, pp. 140--150. [Online]. Available:
  \url{https://doi.org/10.1145/1273463.1273483}
\BIBentrySTDinterwordspacing

\bibitem[Yu and Pradel(2018)]{yu:18}
\BIBentryALTinterwordspacing
T.~Yu and M.~Pradel, ``Pinpointing and repairing performance bottlenecks in
  concurrent programs,'' \emph{Empirical Software Engineering}, vol.~23, no.~5,
  pp. 3034--3071, Nov. 2018. [Online]. Available:
  \url{https://doi.org/10.1007/s10664-017-9578-1}
\BIBentrySTDinterwordspacing

\bibitem[Zhang et~al.(2020)Zhang, Zhang, Yue, Ali, and
  Li]{zhang2020uncertainty}
\BIBentryALTinterwordspacing
H.~Zhang, M.~Zhang, T.~Yue, S.~Ali, and Y.~Li, ``Uncertainty-wise requirements
  prioritization with search,'' \emph{{ACM} Transactions on Software
  Engineering and Methodology}, vol.~30, no.~1, 2020. [Online]. Available:
  \url{https://doi.org/10.1145/3408301}
\BIBentrySTDinterwordspacing

\bibitem[Zitzler and K{\"u}nzli(2004)]{zitzler:04}
\BIBentryALTinterwordspacing
E.~Zitzler and S.~K{\"u}nzli, ``Indicator-based selection in multiobjective
  search,'' in \emph{Proceedings of the 8th International Conference on
  Parallel Problem Solving from Nature}, ser. {PPSN} 2004, vol. 3242.\hskip 1em
  plus 0.5em minus 0.4em\relax Springer, 2004, pp. 832--842. [Online].
  Available: \url{https://doi.org/10.1007/978-3-540-30217-9_84}
\BIBentrySTDinterwordspacing

\bibitem[Zitzler et~al.(2001)Zitzler, Laumanns, and Thiele]{zitzler:01}
\BIBentryALTinterwordspacing
E.~Zitzler, M.~Laumanns, and L.~Thiele, ``Spea2: {I}mproving the strength
  pareto evolutionary algorithm,'' {ETH} Zurich, Computer Engineering and
  Networks Laboratory, {TIK} Report 103, May 2001. [Online]. Available:
  \url{https://doi.org/10.3929/ETHZ-A-004284029}
\BIBentrySTDinterwordspacing

\end{thebibliography}

\end{document}